\documentclass[aip,
 amsmath,amssymb,
 reprint,%
]{revtex4-1}

\usepackage{graphicx}
\usepackage{dcolumn}
\usepackage{bm}

\usepackage[utf8]{inputenc}
\usepackage[T1]{fontenc}
\usepackage{mathptmx}
\usepackage{etoolbox}
\usepackage{color}
\usepackage{hyperref}
\usepackage{array}

\makeatletter
\def\@email#1#2{%
 \endgroup
 \patchcmd{\titleblock@produce}
  {\frontmatter@RRAPformat}
  {\frontmatter@RRAPformat{\produce@RRAP{*#1\href{mailto:#2}{#2}}}\frontmatter@RRAPformat}
  {}{}
}%
\makeatother

\begin{document}
\preprint{AIP/123-QED}

\title{The quantromon: A qubit-resonator system with orthogonal qubit and readout modes}

\author{Kishor V. Salunkhe}
\affiliation{Department of Condensed Matter Physics and Materials Sciences \newline Tata Institute of Fundamental Research, Homi Bhabha Road, Mumbai 400005, India }

\author{Suman Kundu}%
\affiliation{School of Science, Aalto University, P.O. Box 15100, FI-00076 Aalto, Finland.}

\author{Srijita Das}
\affiliation{Department of Condensed Matter Physics and Materials Sciences \newline Tata Institute of Fundamental Research, Homi Bhabha Road, Mumbai 400005, India }

\author{Jay Deshmukh}
\affiliation{Department of Condensed Matter Physics and Materials Sciences \newline Tata Institute of Fundamental Research, Homi Bhabha Road, Mumbai 400005, India }

\author{Meghan P. Patankar}
\affiliation{Department of Condensed Matter Physics and Materials Sciences \newline Tata Institute of Fundamental Research, Homi Bhabha Road, Mumbai 400005, India }

\author{R. Vijay}
\affiliation{Department of Condensed Matter Physics and Materials Sciences \newline Tata Institute of Fundamental Research, Homi Bhabha Road, Mumbai 400005, India }

\date{\today}

\begin{abstract}
The measurement of a superconducting qubit is implemented by coupling it to a resonator. The common choice is transverse coupling, which, in the dispersive approximation, introduces an interaction term which enables the measurement. This cross-Kerr term provides a qubit-state dependent dispersive shift in the resonator frequency with the device parameters chosen carefully to get sufficient signal while minimizing Purcell decay of the qubit. We introduce a two-mode circuit, nicknamed quantromon, with two orthogonal modes implementing a qubit and a resonator. Unlike before, where the coupling term emerges as a perturbative expansion, the quantromon has intrinsic cross-Kerr coupling by design. Our experiments implemented in a hybrid 2D-3D cQED architecture demonstrate some unique features of the quantromon like weak dependence of the dispersive shift on the qubit-resonator detuning and intrinsic Purcell protection. In a tunable qubit-frequency device, we show that the dispersive shift ($2\chi/2\pi$) changes by only $0.8$~MHz while the qubit-resonator detuning ($\Delta/2\pi$) is varied between $0.398$~GHz - $3.288$~GHz. We also demonstrate Purcell protection in a second device where we tune the orthogonality between the two modes. Finally, we demonstrate a single-shot readout fidelity of $98.3\%$ without using a parametric amplifier which is comparable to the state-of-the-art and suggests a potential simplification of the measurement circuitry for scaling up quantum processors.
\end{abstract}
\maketitle
\newpage

Quantum processors using superconducting qubits are in the NISQ (Noisy Intermediate Scale Quantum) era \cite{preskill_quantum_2018} with further demands to improve speed and fidelity of quantum gates and qubit-state measurement towards the ultimate goal of fault tolerant quantum computing. A fast high-fidelity measurement is as crucial as gate fidelities since it is essential for state initialization \cite{sunada_fast_2022,zhou_rapid_2021}, error mitigation\cite{li_efficient_2017}, and more. In circuit QED (cQED), the qubit is coupled to the readout resonator to enable measurement. The widely used choice is transverse or exchange coupling \cite{blais_cavity_2004, wallraff_approaching_2005, boissonneault_dispersive_2009, zueco_qubit-oscillator_2009, jeffrey_fast_2014, walter_rapid_2017} of the form  $g (\hat{a}_{q} + \hat{a}^\dagger_{q})(\hat{a}_r + \hat{a}_r^\dagger)$ where $g$ is the coupling strength and $a_{i} (a^{\dagger}_{i})$ are the annihilation (creation) operators for qubit ($q$) and resonator ($r$).  For a transmon qubit \cite{koch_charge-insensitive_2007} in the dispersive approximation, the interaction term comes out to be $2 \chi \hat{\sigma}_z \hat{a}^\dagger \hat{a}$, where $\chi = ({g^2}/{\Delta})({\alpha_q}/{(\Delta + \alpha_q)})$ is the dispersive shift. Here, $\Delta$ is qubit-resonator detuning and $\alpha_q$ is the transmon's anharmonicity. Hence the coupled resonator has a qubit state-dependent frequency shift ($2 \chi$), and the signal scattered from the resonator reveals the qubit's state. Though high readout fidelities (99.1\%-99.6\% \cite{walter_rapid_2017}) have been reported, one has to work with carefully optimized combinations of $g$, $\Delta$, and $ \kappa$ to achieve sufficient SNR \cite{gambetta_quantum_2008} while avoiding qubit decay via the Purcell effect\cite{e_m_purcell_spontaneous_1946} by using additional circuitry like a  Purcell filter\cite{walter_rapid_2017, reed_fast_2010, sunada_fast_2022}  or tunable coupler\cite{gard_fast_2019} between the qubit and resonator.

Alternative coupling schemes have been explored and readout schemes based on longitudinal coupling\cite{didier_fast_2015} and conditional displacement of the resonator state\cite{touzard_gated_2019} have shown promising results for fast, high-fidelity readout. Multi-modal devices also offer an alternative coupling mechanism between different modes in a single device and have been used in a variety of applications in cQED. The Josephson ring modulator (JRM)\cite{bergeal_analog_2010} can be operated as an amplifier or a frequency converter and relies on the three-way coupling between three modes in a four junction circuit. A three-qubit device with pair-wise cross-Kerr coupling was realized using the multi-modal circuit Trimon \cite{roy_implementation_2017}. More recently, a quantum circuit with control and readout operating modes that successfully decouples the qubit during readout\cite{pfeiffer_efficient_2024} was implemented using a multi-modal circuit.

Here,  we introduce an integrated qubit-resonator system using a multi-modal circuit nicknamed quantromon. The orthogonal modes of the quantromon, a transmon-like qubit and lumped LC-oscillator are coupled via cross-Kerr coupling. Our circuit is an evolution of the quantronium\cite{vijay_invited_2009} circuit where a charge qubit was cross-Kerr coupled to a nonlinear resonator, and hence we named it quantromon. In another implementation of a similar circuit, a two-mode circuit with cross-Kerr coupling between a qubit and an ancilla mode is transversely coupled to the readout resonator via the ancilla \cite{dassonneville_fast_2020}. This creates an effective coupling between the qubit and the readout resonator while retaining their orthogonality and also providing a way to measure the qubit state\cite{roy_implementation_2017}. In this article, we use three different samples to demonstrate the key properties of the quantromon viz. weak dependence of the dispersive shift on qubit-resonator detuning, intrinsic protection from Purcell decay and high readout-fidelity without using Josephson Parametric Amplifiers (JPAs).

\begin{figure}[h!]
\centering
\includegraphics[scale = 0.45]{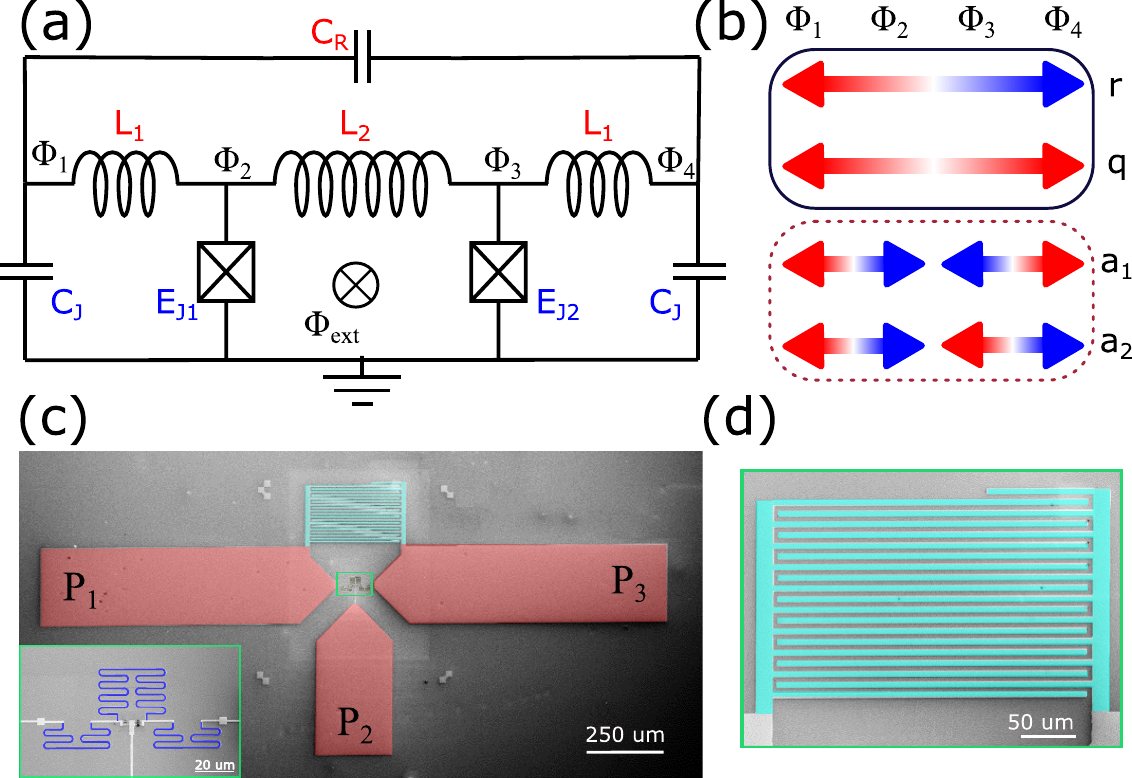}
\caption{(a) Quantromon circuit schematic. (b) The multi-modal circuit has two modes: a transmon-like mode $q$ and a linear resonator mode $r$. The modes $a_1$ and $a_2$ can arise due to the stray capacitance in the system. (c) SEM image (False colored): Three superconducting pads are shown in red. The transmon mode is an effective dipole between P2 and a combination of P1 and P3. The pads P1 and P3 together act as an antenna for the linear mode to set the coupling with the waveguide. In the inset, we show the meander inductor made using a thin line of $500$~nm, split into three parts (blue) as the schematic suggests. The central part $L_2$ forms a loop with two squid Josephson junctions. (d) Interdigital capacitor for the lumped LC oscillator (cyan). The capacitor fingers are $5$~$\mu$m wide and a $5$~$\mu$m gap is maintained between them.}
\label{Schematic and SEM Image}
\end{figure}

The SEM image of the quantromon is shown in Fig. \ref{Schematic and SEM Image}(c), which implements the lumped circuit shown in Fig. \ref{Schematic and SEM Image}(a). The capacitance $C_R$ is realized by using an interdigital capacitor (Fig. \ref{Schematic and SEM Image}(d)) with $5$~$\mu$m width and $5$~$\mu$m gap. The linear inductor is made by using a $500$~nm wide meandering line (Fig. \ref{Schematic and SEM Image}(c) inset). The inductance $L_R$ is divided into three sections, of which two are of equal length ($L_1 = (1-b) L_R/2$) and the middle section ($L_2 = b L_R$) is part of the quantromon loop formed with two Josephson junctions with Josephson energies $E_{J1}$ and $E_{J2}$. The ratio $b=L_2/L_R \approx 0.4$ for the devices used in this paper and controls the effective coupling between the qubit and the resonator modes. The capacitance $C_J$ across the junctions is due to the proximity of superconducting pads P1 and P3 with the center pad P2. Pads P1 and P3 together act as an antenna for the readout resonator so that it can be excited by a 3D rectangular waveguide (see supplementary material). This is a hybrid 2D-3D cQED architecture where the qubit mode behaves like a 3D transmon while the readout mode is a lumped LC oscillator coupled to a rectangular 3D waveguide for control and measurement \cite{kundu_multiplexed_2019}.

This multi-modal circuit is analyzed at zero flux using four differential modes as shown in Fig.\ref{Schematic and SEM Image} (b) where we have assumed the ideal case with $E_{J1}=E_{J2}=E_{J}$ i.e. zero asymmetry in the qubit junctions. The modes relevant to us are the mode $q$, where the charge oscillates between the equipotential nodes $\Phi_1$ and $\Phi_4$, and the ground node, and the mode $r$, where the charge oscillates between nodes $\Phi_1$ and $\Phi_4$. Here we ignore the high-frequency modes $a_1$ and $a_2$, which may arise due to the stray capacitance (not shown) from node $\Phi_1$ to $\Phi_2$ and $\Phi_3$ to $\Phi_4$. The Hamiltonian of the system in the transformed variables $\phi_q = (\Phi_1 + \Phi_4)/2\Phi_o$ and $\phi_r = (\Phi_1 - \Phi_4)/\Phi_o$ is given by,

\begin{align}\label{Eq.1}
H &= E_{{c}_{Q}}\hat{q}_q^2 + E_{{c}_{R}} \hat{q}_r^2 + \frac{E_{{J}_{q}}}{2} \hat{\phi}_q^2 - \frac{E_J}{12}\hat{\phi}_q^4 \notag \\
&\quad + \frac{E_{{J}_{r}}}{2} \hat{\phi}_r^2 - \frac{b^4}{192}E_J\hat{\phi}_r^4 - \frac{b^2}{8}E_J\hat{\phi}_q^2\hat{\phi}_r^2
\end{align}
where $E_{{C}_{Q}} = e^2 /(4 C_J)$ and $E_{{C}_{R}} = e^2/(C_J + 2 C_R) $ are the charging energies of the qubit and resonator modes respectively. $E_{{J}_{q}} = 2 E_J$ and $E_{{J}_{r}} = E_{L_R} + (b^2/2)E_J$ are the inductive energies for the qubit and resonator modes respectively, where $E_{L_R} = \phi_o^2/ L_R$. The last term is the direct cross-Kerr coupling term between the qubit and resonator mode. The device is operated at integer flux quantum in the quantromon loop to suppress three-wave mixing terms \cite{bergeal_analog_2010,roy_implementation_2017} and dephasing due to flux noise.  The Hamiltonian in the second quantised form can now be written as,

\begin{equation}
\frac{H}{\hbar} = \sum_{i = q,r} (\omega_i - J_i - J_{qr}) \hat{n}_i - J_i \hat{n}_i^2 - 2\chi\hat{n}_q\hat{n}_r
\end{equation}
where $\omega_{i = q,r}$ and $2J_{i =q,r}$ are the uncoupled frequencies and anharmonicities of the qubit and resonator modes respectively. We analyze this circuit in the limit $E_{L_R} >> E_J$ which implies that the nonlinearity of the resonator mode is extremely weak and can be ignored. The dispersive shift is given by

\begin{equation}
    2 \chi = \frac{b^2}{\sqrt{2}}\sqrt{\frac{E_{C_Q} E_{C_R}}{\frac{b^2}{2} + \frac{E_{L_R}}{E_J}}}
\end{equation}
and weakly depends upon the qubit-resonator detuning via the ratio $E_{L_R}/E_J$ (see supplementary material). This can enable large dispersive shifts and a wider choice of qubit-resonator detuning without suffering from Purcell decay since the qubit and resonator modes are orthogonal to each other by design (assuming zero or small asymmetry i.e. $E_{J1} \approx E_{J2}$)

\begin{figure}[h!]
\centering
\includegraphics[scale = 0.5]{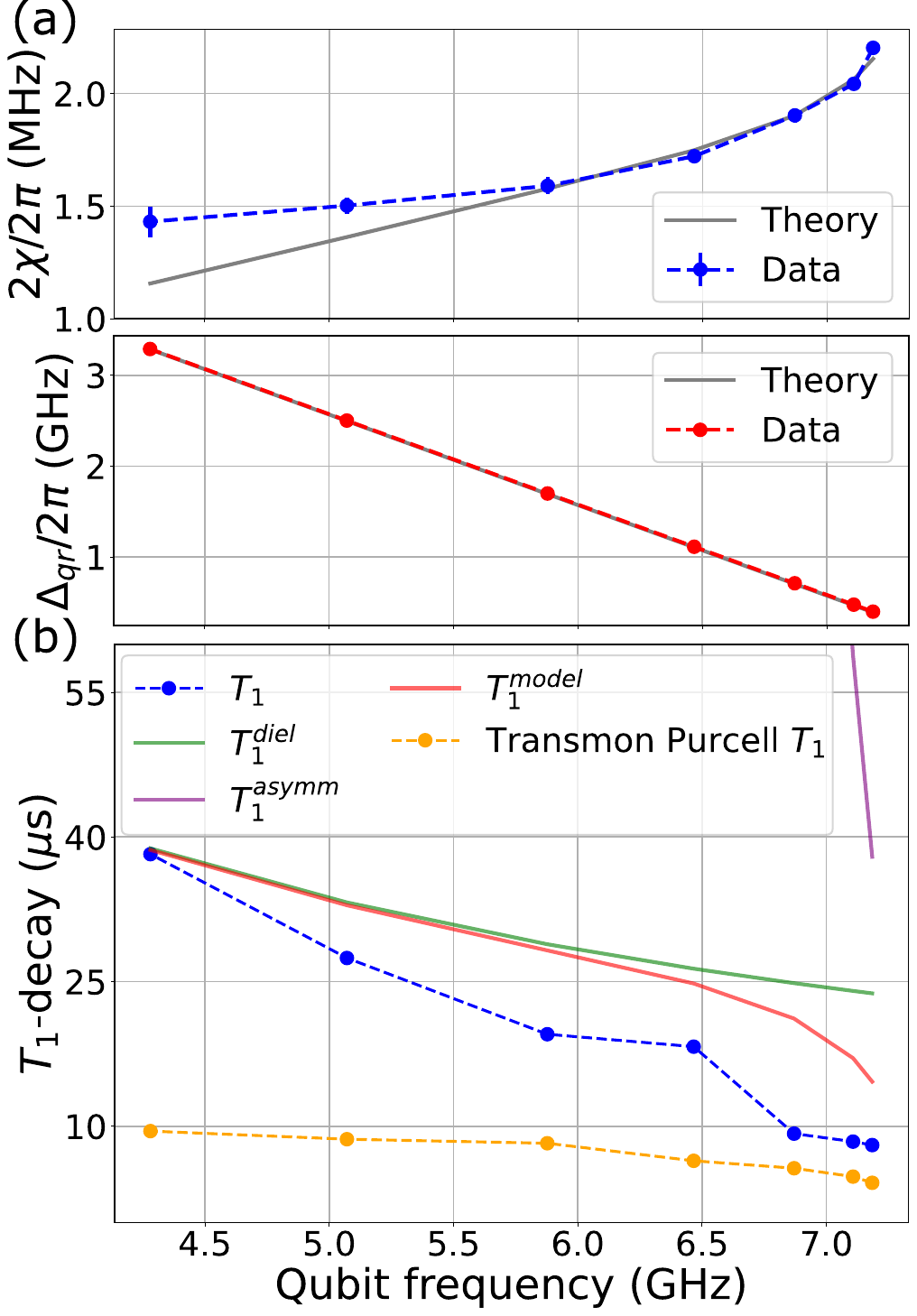}
\caption{(a) The dispersive shift (blue) and qubit-resonator detuning (red) at different qubit frequencies. This shows the weak dependence of dispersive shift on detuning as it changes from $2.2$~MHz to $1.4$~MHz as detuning changes from $0.398$~GHz to $3.288$~GHz (b) The quantromon experimental $T_1$ decay time (blue) is more than the Purcell $T_1$ (yellow) of the equivalent transmon. The decreasing trend in the $T_1$ is explained by the dielectric loss and the Purcell effect due to the asymmetry.}
\label{Dispersive shift}
\end{figure}

The three samples used in the main results of the paper are fabricated using two different methods and measured using a rectangular waveguide (see supplementary material for details). In Sample A, the individual Josephson junctions in the non-tunable quantromon (Fig. \ref{Schematic and SEM Image}(a)) are replaced by SQUIDs so that we can tune $E_{J_q}=E_{J1}+E_{J2}$. The loop area ratio of the SQUID to the quantromon loop is 6.8\%. Since we can only operate at integer flux quanta in the quantromon loop,  we can tune the qubit frequency in discrete steps and study the changes in the dispersive shift as a function of qubit-resonator detuning. Note that the readout resonator frequency is very weakly dependent on the value of $E_{J_q}$ as $E_{{J}_{r}} = E_{L_R} + (b^2/2)E_J$ and we operate in the regime $E_{L_R} >> E_J$. We measure the dispersive shift ($2\chi/2\pi$)  at different qubit frequencies varying from $7.185$ GHz (0-flux quantum) to $4.281$ GHz (9-flux quantum) and are plotted in Fig. \ref{Dispersive shift}(a). The dispersive shift changes from $2.2$~MHz to $1.4$~MHz while the qubit-resonator ($\Delta_{qr}/2\pi$) detuning changes from $0.398$~GHz to $3.288$~GHz. As the detuning increases linearly with qubit frequency (right to left), the dispersive shift slowly decreases. The asymmetry between the two SQUIDs, $d_j = ({E_{J1} - E_{J2}})/({E_{J1} + E_{J2}})$ can explain the faster changes at smaller detunings as even small asymmetry-induced transverse coupling can produce large dispersive shifts. A theory prediction with $4.5\%$ junction asymmetry shows good agreement with experiment at smaller detunings. The small deviation at larger detunings is likely due to the asymmetry changing a lot as the two SQUIDs approach half flux quantum in their respective loops due to the finite asymmetry between the junctions of each SQUID. Nevertheless, the data clearly shows that the dispersive shift in the quantromon has a weak dependence on the qubit-resonator detuning.

The $T_1$ values at all the operating points are plotted in blue (Fig.\ref{Dispersive shift}(b)) and shows a steady decrease with increasing qubit frequency. Using $4.5\%$ asymmetry we compute the $T_1^{asymm}$ (due to the Purcell effect; see supplementary material) and is shown by a solid purple line (visible only on the top right part of the graph). Due to the orthogonality of the qubit and resonator mode and the small asymmetry between the two qubit SQUIDs, the Purcell effect starts to affect the $T_1^{asymm}$ only at small detunings. We then fix the dielectric loss to match the data at the lowest qubit frequency since the Purcell effect can be ignored at this operating point. The Purcell decay along with the dielectric loss ($T_1^{diel}$) with a quality factor of $1.1\times 10^6$ predicts the trend observed in the data. The prediction ${1}/{T_1^{model}} = {1}/{T_1^{diel}} + {1}/{T_1^{asymm}}$ is shown by solid red line and follows the trend shown by the experimental data. The disagreement is likely due to enhanced losses at higher qubit frequencies, a trend we typically observe in all our devices. Finally, we also compare the quantromon $T_1$ with that of a transmon coupled to a readout resonator and providing the same dispersive shifts with no Purcell filter. We calculate the coupling required to achieve the measured dispersive shift in the quantromon to estimate the Purcell $T_1$ of the transversely coupled transmon. The transmon Purcell $T_1$ is plotted in yellow and is much lower than the experimentally measured $T_1$ values of the quantromon. This demonstrates the possibility of obtaining strong dispersive shifts at large detunings with protection from Purcell decay.

\begin{figure}[h!]
\centering
\includegraphics[scale = 0.5]{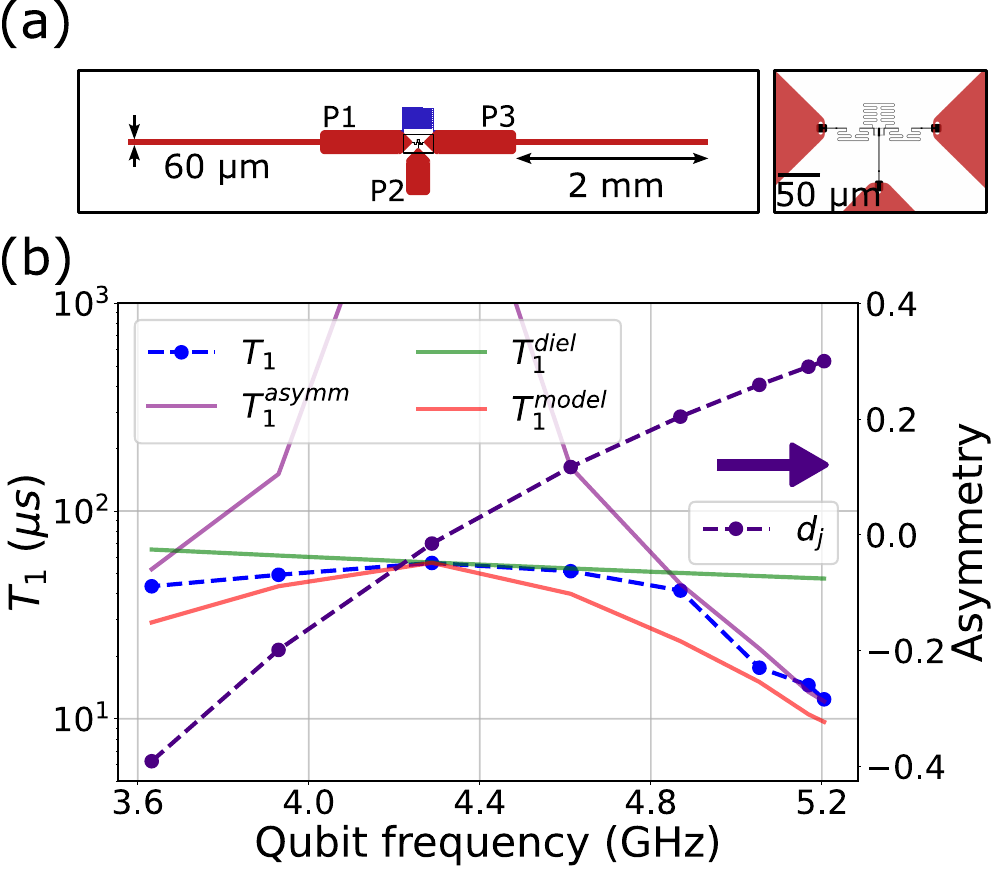}
\caption{Purcell Protection: (a) A schematic of the tunable asymmetry quantromon. The antennas are modified by adding $2$~mm long and $60$~$\mu$m wide to increase the coupling to the environment. The modification is far from the center pad and keeps the qubit capacitance unaffected. A notch-type structure is used to ensure the galvanic contact of the tantalum (brown) and aluminum(black). (b) The measured $T_1$-decay numbers are shown in blue. The extracted asymmetry variation is shown on the right axis. The model $T_1^{model}$ combines the dielectric loss ($T_1^{diel}$) and Purcell effect due to asymmetry ($T_1^{asymm}$).}
\label{Purcell Protection}
\end{figure}

In Sample A, the Purcell decay due to asymmetry-induced transverse coupling is relevant only for small detunings. To further explore the Purcell effect due to asymmetry in a quantromon, we designed Sample B where only one of the Josephson junctions in Fig. \ref{Schematic and SEM Image}(a) is replaced by a SQUID of area $A_{SQUID}$. This allows us to tune the asymmetry of the device but the qubit frequency also changes since the effective inductive energy $E_{J\Sigma} = E_{J1} + E_{J2}\cos(n \pi a)$, where $a = {A_{SQUID}}/{A_{Quantro}}$, changes as we change the flux quanta $n$. We also increased the coupling of the resonator mode with the environment to enhance the Purcell effect due to asymmetry and to clearly demonstrate the protection at operating points with small asymmetries. As shown in Fig.\ref{Purcell Protection}(a), a $2$ mm long and $60$ $\mu$m wide section is added to the pads P1 and P3 to increase the linewidth of the resonator by $\sim 20$ times. The $E_J$ of the SQUID was made to be much larger than $E_J$ of the single junction so that we start with a highly asymmetric device at zero flux quantum and slowly decrease the asymmetry as the $E_J$ of the SQUID tunes down with flux.

Like the previous experiment, we operated the device at multiple integer flux quanta in the quantromon loop and measured the relaxation times. The device has $30\% $ asymmetry at zero flux ($5.205$ GHz) as estimated from the room temperature resistance measurements. We then estimate the asymmetry at all other operating points by varying $E_J$ of the SQUID so that the estimated qubit frequency matches with the measured value at that operating point. The extracted asymmetry variation is plotted on the right axis in Fig. \ref{Purcell Protection}(b). The correlation between the strength of the asymmetry and experimentally measured $T_1$ (Fig.\ref{Purcell Protection}(b)) values can be seen. The +5-flux quanta operation point ($4.288$ GHz)has the minimum asymmetry $-1.52\%$ and the measured $T_1$ is the maximum at this bias point. The $T_1$ behavior can be explained by $T_1^{asymm}$ and $T_1^{diel}$ shown by the solid purple and green line respectively in Fig.\ref{Purcell Protection}(b). For this device, $T_1^{diel}$ is computed using a quality factor $1.61\times10^6$ and the model used qualitatively captures the trend observed in the data. This demonstrates the in-built Purcell protection of the quantromon along with sufficient tolerance for small asymmetries which are invariably present due to fabrication variabilities. This evades the need for additional circuitry for Purcell filters \cite{walter_rapid_2017, reed_fast_2010, sunada_fast_2022} that are typically used to engineer the qubit environment in a conventional transmon coupled dispersively to a readout resonator.

\begin{figure}[h]
\centering
\includegraphics[scale = 0.22]{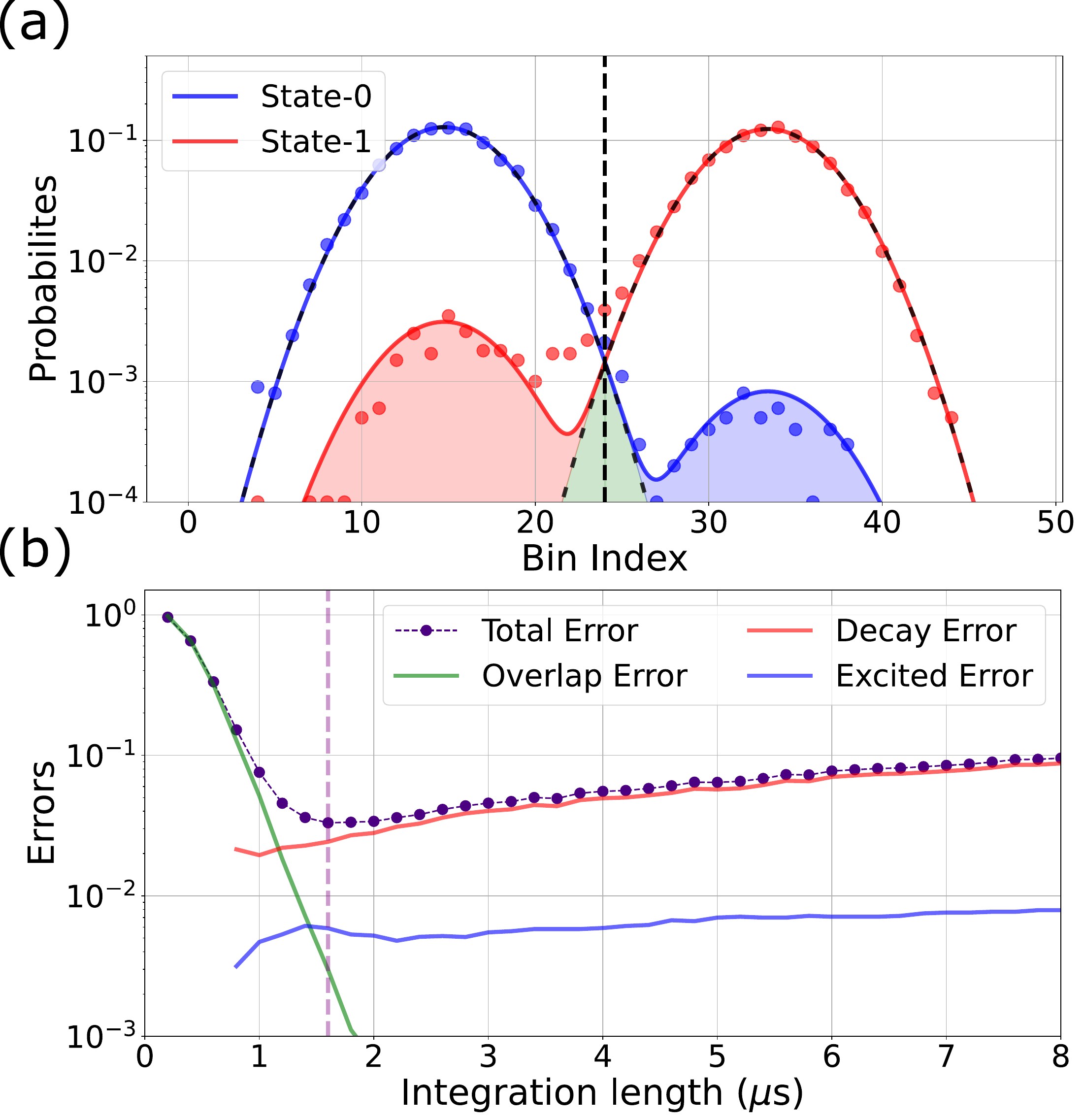}
\caption{(a) Histogram of 1.8 $\mu$s integration length for qubit prepared in the state-0 (blue) and state-1 (red). We fit the double Gaussian model (solid blue and red lines) to the data. The green area shows the overlap error ($\epsilon_{id}$). The blue and red areas show the other error $\epsilon_{01}$ and $\epsilon_{10}$ respectively (b) The error analysis with integration time does show that overlap error (dashed green line) is significant at low integration times and decay error (dashed red line) dominates at large integration times. The excited error is below $1\%$ at all times. The point with maximum readout fidelity (histograms shown in (a)) is marked with the vertical purple line.}
\label{Histograms}
\end{figure}

Finally, we discuss experiments on Sample C, a non-tunable quantromon designed to study qubit readout at high photon numbers in the resonator without using a JPA. We extract qubit readout fidelity (also referred to as state assignment fidelity \cite{chen_transmon_2023}) by measuring single-shot histograms with the qubit prepared in the ground and excited states as shown in Fig. \ref{Histograms}(a). The single-shot histograms are fitted with bimodal Gaussian distributions  $A_0 N(x, \mu_0, \sigma_0) + (1 - A_0)N(x, \mu_1, \sigma_1)$ for state-0 and $(1 - A_1) N(x, \mu_0, \sigma_0) + A_1N(x, \mu_1, \sigma_1)$ for state-1 simultaneously to extract $\mu_0$, $\mu_1$, $\sigma_0$ and $\sigma_1$\cite{walter_rapid_2017}. The fitted double Gaussian distributions are plotted in solid colors and the single Gaussians $A_0 N(x, \mu_0, \sigma_0)$ and $A_1N(x, \mu_1, \sigma_1)$ with dashed black lines. We set the demarcation to be the intersection of the $N(x, \mu_0, \sigma_0)$ and $N(x, \mu_1, \sigma_1)$ to calculate the probability $P(A|B)$ for the qubit to be measured in state-A when prepared in state-B. The readout fidelity is then given by
\begin{equation}
    F = 1 - \frac{P(0|1) + P(1|0)}{2}.
\end{equation}
\par
In Fig. \ref{Histograms}(a), we show the histograms for the best readout fidelity of $98.3\%$ we obtained for this sample. The data shown corresponds to a photon number $\bar{n}\approx30$ in the readout resonator and an integration time of $1.8$ $\mu$s. The dispersive shift $2\chi/2\pi=1.37$ MHz, and the readout resonator linewidth $\kappa/2\pi=1.28$ MHz with $\kappa_{ext}/2\pi=0.90$ MHz and $\kappa_{int}/2\pi=0.38$ MHz. The readout frequency was chosen to be the mean of the resonator frequencies corresponding to the qubit state-0 and state-1, resulting in a reflected signal phase shift of $236 ^{\circ}$. The large photon number combined with the large phase shift is clearly reflected in the small error $\epsilon_{id} = 0.3\%$ calculated from the overlap of the single Gaussians leading to an ideal readout fidelity \cite{chen_transmon_2023} of $99.85\%$. This number only accounts for the SNR of the histograms and ignores any state preparation of mixing (excitation and relaxation) errors. We extract the combination of state preparation and mixing errors as $\epsilon_{01} = 2.4\%$, $\epsilon_{10} = 0.6\%$ respectively. Since we did not use heralding, we cannot separate state preparation errors from mixing errors which are still quite low for the relatively large photon number used in this experiment.We also plot how these errors change with integration time in Fig. \ref{Histograms}(b). We see that overlap error is dominant for small integration lengths as expected while the error at larger times is predominantly due to $T_1$ relaxation with negligible error due to qubit excitation. This suggests that the quantromon is less susceptible to state mixing and ionization \cite{dumas_measurement-induced_2024} compared to a standard transmon qubit coupled transversely to readout resonators. A very recent theoretical analysis for similar circuits suggests that the non-perturbative nature of the cross-Kerr term in such circuits may be responsible for suppressing qubit ionization \cite{chapple_robustness_2024,chapple_balanced_2025}.  Experiments on several other devices yielded high fidelity numbers (see supplementary material) suggesting that this effect is robust.

In conclusion, we have presented a qubit-resonator system with orthogonal qubit and readout modes and demonstrated three key properties. We show that the device has intrinsic protection from Purcell decay, a weak dependence of dispersive shift on qubit-resonator detuning and demonstrated high readout fidelity ($>97.5\%$) without using a JPA. The quasi-constant dispersive shift can offer more flexibility in choosing the resonator frequency and thus giving larger frequency space for multiplexing qubit readout. The symmetry based intrinsic Purcell protection implies that the dispersive shift can be made larger for faster readout but without requiring Purcell filters, which simplifies the circuitry. The quantromon does need a flux bias to operate at integer flux quantum, but tunable qubits are becoming more common and may provide more options for two-qubit gates. The high readout fidelity without using a JPA suggests the ability to operate at high photon numbers without suffering from qubit ionization. The combination of the three key properties demonstrated in our experiment for the quantromon device may help simplify the architecture of future medium to large-scale processors.

\textit{Note.-} During the preparation of this manuscript, we learned of an experimental demonstration of a similar device with non-perturbative cross-Kerr term but one which is optimized to implement a non-linear readout resonator and uses two-state readout based on bifurcation physics \cite{wang_999-fidelity_2024}

See supplementary material for additional theoretical analysis, experimental data and measurement setup details.

\textit{Acknowledgments:-} We acknowledge the nanofabrication and central workshop facilities of TIFR. This work is supported by the Department of Atomic Energy of the Government of India under Project No. RTI4003. We also acknowledge support from the Department of Science and Technology, India, via the QuEST program.

\section*{Author Declarations}

\subsection*{Conflict of Interest}
The authors have no conflicts of interest to disclose.

\subsection*{Authors Contribution}
\noindent \textbf{Kishor V Salunkhe}: Conceptualization (equal); Data curation (lead); Formal analysis (lead);  Investigation (equal); Methodology (equal); Software (equal); Writing -- original draft (lead); Writing -- review and editing (equal).
\textbf{Suman Kundu}: Conceptualization (supporting); Data curation (supporting); Methodology (supporting); Writing -- review and editing (supporting).
\textbf{Srijita Das}: Investigation (supporting); Methodology (supporting); Writing -- review and editing (supporting).
\textbf{Jay Deshmukh}: Investigation (supporting); Software (equal); Writing -- review and editing (supporting).
\textbf{Meghan P Patankar}: Methodology (supporting).
\textbf{R Vijay}: Conceptualization (equal); Investigation (equal); Supervision (lead); Writing -- original draft (supporting); Writing -- review and editing (equal).

\textit{Data Availability:-} The experimental data used in the main paper figures, along with instructions to recreate the plots, are available in Zenodo with the identifier \url{https://doi.org/10.5281/zenodo.14634897}. The raw data and the analysis codes related to this study are available from the corresponding authors upon reasonable request.

\bibliography{Quantromon_reference_v6}

\begin{thebibliography}{28}%
\makeatletter
\providecommand \@ifxundefined [1]{%
 \@ifx{#1\undefined}
}%
\providecommand \@ifnum [1]{%
 \ifnum #1\expandafter \@firstoftwo
 \else \expandafter \@secondoftwo
 \fi
}%
\providecommand \@ifx [1]{%
 \ifx #1\expandafter \@firstoftwo
 \else \expandafter \@secondoftwo
 \fi
}%
\providecommand \natexlab [1]{#1}%
\providecommand \enquote  [1]{``#1''}%
\providecommand \bibnamefont  [1]{#1}%
\providecommand \bibfnamefont [1]{#1}%
\providecommand \citenamefont [1]{#1}%
\providecommand \href@noop [0]{\@secondoftwo}%
\providecommand \href [0]{\begingroup \@sanitize@url \@href}%
\providecommand \@href[1]{\@@startlink{#1}\@@href}%
\providecommand \@@href[1]{\endgroup#1\@@endlink}%
\providecommand \@sanitize@url [0]{\catcode `\\12\catcode `\$12\catcode `\&12\catcode `\#12\catcode `\^12\catcode `\_12\catcode `\%12\relax}%
\providecommand \@@startlink[1]{}%
\providecommand \@@endlink[0]{}%
\providecommand \url  [0]{\begingroup\@sanitize@url \@url }%
\providecommand \@url [1]{\endgroup\@href {#1}{\urlprefix }}%
\providecommand \urlprefix  [0]{URL }%
\providecommand \Eprint [0]{\href }%
\providecommand \doibase [0]{http://dx.doi.org/}%
\providecommand \selectlanguage [0]{\@gobble}%
\providecommand \bibinfo  [0]{\@secondoftwo}%
\providecommand \bibfield  [0]{\@secondoftwo}%
\providecommand \translation [1]{[#1]}%
\providecommand \BibitemOpen [0]{}%
\providecommand \bibitemStop [0]{}%
\providecommand \bibitemNoStop [0]{.\EOS\space}%
\providecommand \EOS [0]{\spacefactor3000\relax}%
\providecommand \BibitemShut  [1]{\csname bibitem#1\endcsname}%
\let\auto@bib@innerbib\@empty
\bibitem [{\citenamefont {Preskill}(2018)}]{preskill_quantum_2018}%
  \BibitemOpen
  \bibfield  {author} {\bibinfo {author} {\bibfnamefont {J.}~\bibnamefont {Preskill}},\ }\bibfield  {title} {\enquote {\bibinfo {title} {Quantum {Computing} in the {NISQ} era and beyond},}\ }\href {\doibase 10.22331/q-2018-08-06-79} {\bibfield  {journal} {\bibinfo  {journal} {Quantum}\ }\textbf {\bibinfo {volume} {2}},\ \bibinfo {pages} {79} (\bibinfo {year} {2018})}\BibitemShut {NoStop}%
\bibitem [{\citenamefont {Sunada}\ \emph {et~al.}(2022)\citenamefont {Sunada}, \citenamefont {Kono}, \citenamefont {Ilves}, \citenamefont {Tamate}, \citenamefont {Sugiyama}, \citenamefont {Tabuchi},\ and\ \citenamefont {Nakamura}}]{sunada_fast_2022}%
  \BibitemOpen
  \bibfield  {author} {\bibinfo {author} {\bibfnamefont {Y.}~\bibnamefont {Sunada}}, \bibinfo {author} {\bibfnamefont {S.}~\bibnamefont {Kono}}, \bibinfo {author} {\bibfnamefont {J.}~\bibnamefont {Ilves}}, \bibinfo {author} {\bibfnamefont {S.}~\bibnamefont {Tamate}}, \bibinfo {author} {\bibfnamefont {T.}~\bibnamefont {Sugiyama}}, \bibinfo {author} {\bibfnamefont {Y.}~\bibnamefont {Tabuchi}}, \ and\ \bibinfo {author} {\bibfnamefont {Y.}~\bibnamefont {Nakamura}},\ }\bibfield  {title} {\enquote {\bibinfo {title} {Fast {Readout} and {Reset} of a {Superconducting} {Qubit} {Coupled} to a {Resonator} with an {Intrinsic} {Purcell} {Filter}},}\ }\href {\doibase 10.1103/PhysRevApplied.17.044016} {\bibfield  {journal} {\bibinfo  {journal} {Phys. Rev. Applied}\ }\textbf {\bibinfo {volume} {17}},\ \bibinfo {pages} {044016} (\bibinfo {year} {2022})}\BibitemShut {NoStop}%
\bibitem [{\citenamefont {Zhou}\ \emph {et~al.}(2021)\citenamefont {Zhou}, \citenamefont {Zhang}, \citenamefont {Yin}, \citenamefont {Huai}, \citenamefont {Gu}, \citenamefont {Xu}, \citenamefont {Allcock}, \citenamefont {Liu}, \citenamefont {Xi}, \citenamefont {Yu}, \citenamefont {Zhang}, \citenamefont {Zhang}, \citenamefont {Li}, \citenamefont {Song}, \citenamefont {Wang}, \citenamefont {Zheng}, \citenamefont {An}, \citenamefont {Zheng},\ and\ \citenamefont {Zhang}}]{zhou_rapid_2021}%
  \BibitemOpen
  \bibfield  {author} {\bibinfo {author} {\bibfnamefont {Y.}~\bibnamefont {Zhou}}, \bibinfo {author} {\bibfnamefont {Z.}~\bibnamefont {Zhang}}, \bibinfo {author} {\bibfnamefont {Z.}~\bibnamefont {Yin}}, \bibinfo {author} {\bibfnamefont {S.}~\bibnamefont {Huai}}, \bibinfo {author} {\bibfnamefont {X.}~\bibnamefont {Gu}}, \bibinfo {author} {\bibfnamefont {X.}~\bibnamefont {Xu}}, \bibinfo {author} {\bibfnamefont {J.}~\bibnamefont {Allcock}}, \bibinfo {author} {\bibfnamefont {F.}~\bibnamefont {Liu}}, \bibinfo {author} {\bibfnamefont {G.}~\bibnamefont {Xi}}, \bibinfo {author} {\bibfnamefont {Q.}~\bibnamefont {Yu}}, \bibinfo {author} {\bibfnamefont {H.}~\bibnamefont {Zhang}}, \bibinfo {author} {\bibfnamefont {M.}~\bibnamefont {Zhang}}, \bibinfo {author} {\bibfnamefont {H.}~\bibnamefont {Li}}, \bibinfo {author} {\bibfnamefont {X.}~\bibnamefont {Song}}, \bibinfo {author} {\bibfnamefont {Z.}~\bibnamefont {Wang}}, \bibinfo {author} {\bibfnamefont {D.}~\bibnamefont {Zheng}}, \bibinfo {author} {\bibfnamefont
  {S.}~\bibnamefont {An}}, \bibinfo {author} {\bibfnamefont {Y.}~\bibnamefont {Zheng}}, \ and\ \bibinfo {author} {\bibfnamefont {S.}~\bibnamefont {Zhang}},\ }\bibfield  {title} {\enquote {\bibinfo {title} {Rapid and unconditional parametric reset protocol for tunable superconducting qubits},}\ }\href {\doibase 10.1038/s41467-021-26205-y} {\bibfield  {journal} {\bibinfo  {journal} {Nat Commun}\ }\textbf {\bibinfo {volume} {12}},\ \bibinfo {pages} {5924} (\bibinfo {year} {2021})},\ \bibinfo {note} {publisher: Nature Publishing Group}\BibitemShut {NoStop}%
\bibitem [{\citenamefont {Li}\ and\ \citenamefont {Benjamin}(2017)}]{li_efficient_2017}%
  \BibitemOpen
  \bibfield  {author} {\bibinfo {author} {\bibfnamefont {Y.}~\bibnamefont {Li}}\ and\ \bibinfo {author} {\bibfnamefont {S.~C.}\ \bibnamefont {Benjamin}},\ }\bibfield  {title} {\enquote {\bibinfo {title} {Efficient {Variational} {Quantum} {Simulator} {Incorporating} {Active} {Error} {Minimization}},}\ }\href {\doibase 10.1103/PhysRevX.7.021050} {\bibfield  {journal} {\bibinfo  {journal} {Phys. Rev. X}\ }\textbf {\bibinfo {volume} {7}},\ \bibinfo {pages} {021050} (\bibinfo {year} {2017})}\BibitemShut {NoStop}%
\bibitem [{\citenamefont {Blais}\ \emph {et~al.}(2004)\citenamefont {Blais}, \citenamefont {Huang}, \citenamefont {Wallraff}, \citenamefont {Girvin},\ and\ \citenamefont {Schoelkopf}}]{blais_cavity_2004}%
  \BibitemOpen
  \bibfield  {author} {\bibinfo {author} {\bibfnamefont {A.}~\bibnamefont {Blais}}, \bibinfo {author} {\bibfnamefont {R.-S.}\ \bibnamefont {Huang}}, \bibinfo {author} {\bibfnamefont {A.}~\bibnamefont {Wallraff}}, \bibinfo {author} {\bibfnamefont {S.~M.}\ \bibnamefont {Girvin}}, \ and\ \bibinfo {author} {\bibfnamefont {R.~J.}\ \bibnamefont {Schoelkopf}},\ }\bibfield  {title} {\enquote {\bibinfo {title} {Cavity quantum electrodynamics for superconducting electrical circuits: {An} architecture for quantum computation},}\ }\href {\doibase 10.1103/PhysRevA.69.062320} {\bibfield  {journal} {\bibinfo  {journal} {Phys. Rev. A}\ }\textbf {\bibinfo {volume} {69}},\ \bibinfo {pages} {062320} (\bibinfo {year} {2004})}\BibitemShut {NoStop}%
\bibitem [{\citenamefont {Wallraff}\ \emph {et~al.}(2005)\citenamefont {Wallraff}, \citenamefont {Schuster}, \citenamefont {Blais}, \citenamefont {Frunzio}, \citenamefont {Majer}, \citenamefont {Devoret}, \citenamefont {Girvin},\ and\ \citenamefont {Schoelkopf}}]{wallraff_approaching_2005}%
  \BibitemOpen
  \bibfield  {author} {\bibinfo {author} {\bibfnamefont {A.}~\bibnamefont {Wallraff}}, \bibinfo {author} {\bibfnamefont {D.~I.}\ \bibnamefont {Schuster}}, \bibinfo {author} {\bibfnamefont {A.}~\bibnamefont {Blais}}, \bibinfo {author} {\bibfnamefont {L.}~\bibnamefont {Frunzio}}, \bibinfo {author} {\bibfnamefont {J.}~\bibnamefont {Majer}}, \bibinfo {author} {\bibfnamefont {M.~H.}\ \bibnamefont {Devoret}}, \bibinfo {author} {\bibfnamefont {S.~M.}\ \bibnamefont {Girvin}}, \ and\ \bibinfo {author} {\bibfnamefont {R.~J.}\ \bibnamefont {Schoelkopf}},\ }\bibfield  {title} {\enquote {\bibinfo {title} {Approaching {Unit} {Visibility} for {Control} of a {Superconducting} {Qubit} with {Dispersive} {Readout}},}\ }\href {\doibase 10.1103/PhysRevLett.95.060501} {\bibfield  {journal} {\bibinfo  {journal} {Phys. Rev. Lett.}\ }\textbf {\bibinfo {volume} {95}},\ \bibinfo {pages} {060501} (\bibinfo {year} {2005})}\BibitemShut {NoStop}%
\bibitem [{\citenamefont {Boissonneault}, \citenamefont {Gambetta},\ and\ \citenamefont {Blais}(2009)}]{boissonneault_dispersive_2009}%
  \BibitemOpen
  \bibfield  {author} {\bibinfo {author} {\bibfnamefont {M.}~\bibnamefont {Boissonneault}}, \bibinfo {author} {\bibfnamefont {J.~M.}\ \bibnamefont {Gambetta}}, \ and\ \bibinfo {author} {\bibfnamefont {A.}~\bibnamefont {Blais}},\ }\bibfield  {title} {\enquote {\bibinfo {title} {Dispersive regime of circuit {QED}: {Photon}-dependent qubit dephasing and relaxation rates},}\ }\href {\doibase 10.1103/PhysRevA.79.013819} {\bibfield  {journal} {\bibinfo  {journal} {Phys. Rev. A}\ }\textbf {\bibinfo {volume} {79}},\ \bibinfo {pages} {013819} (\bibinfo {year} {2009})}\BibitemShut {NoStop}%
\bibitem [{\citenamefont {Zueco}\ \emph {et~al.}(2009)\citenamefont {Zueco}, \citenamefont {Reuther}, \citenamefont {Kohler},\ and\ \citenamefont {Hänggi}}]{zueco_qubit-oscillator_2009}%
  \BibitemOpen
  \bibfield  {author} {\bibinfo {author} {\bibfnamefont {D.}~\bibnamefont {Zueco}}, \bibinfo {author} {\bibfnamefont {G.~M.}\ \bibnamefont {Reuther}}, \bibinfo {author} {\bibfnamefont {S.}~\bibnamefont {Kohler}}, \ and\ \bibinfo {author} {\bibfnamefont {P.}~\bibnamefont {Hänggi}},\ }\bibfield  {title} {\enquote {\bibinfo {title} {Qubit-oscillator dynamics in the dispersive regime: {Analytical} theory beyond the rotating-wave approximation},}\ }\href {\doibase 10.1103/PhysRevA.80.033846} {\bibfield  {journal} {\bibinfo  {journal} {Phys. Rev. A}\ }\textbf {\bibinfo {volume} {80}},\ \bibinfo {pages} {033846} (\bibinfo {year} {2009})}\BibitemShut {NoStop}%
\bibitem [{\citenamefont {Jeffrey}\ \emph {et~al.}(2014)\citenamefont {Jeffrey}, \citenamefont {Sank}, \citenamefont {Mutus}, \citenamefont {White}, \citenamefont {Kelly}, \citenamefont {Barends}, \citenamefont {Chen}, \citenamefont {Chen}, \citenamefont {Chiaro}, \citenamefont {Dunsworth}, \citenamefont {Megrant}, \citenamefont {O’Malley}, \citenamefont {Neill}, \citenamefont {Roushan}, \citenamefont {Vainsencher}, \citenamefont {Wenner}, \citenamefont {Cleland},\ and\ \citenamefont {Martinis}}]{jeffrey_fast_2014}%
  \BibitemOpen
  \bibfield  {author} {\bibinfo {author} {\bibfnamefont {E.}~\bibnamefont {Jeffrey}}, \bibinfo {author} {\bibfnamefont {D.}~\bibnamefont {Sank}}, \bibinfo {author} {\bibfnamefont {J.~Y.}\ \bibnamefont {Mutus}}, \bibinfo {author} {\bibfnamefont {T.~C.}\ \bibnamefont {White}}, \bibinfo {author} {\bibfnamefont {J.}~\bibnamefont {Kelly}}, \bibinfo {author} {\bibfnamefont {R.}~\bibnamefont {Barends}}, \bibinfo {author} {\bibfnamefont {Y.}~\bibnamefont {Chen}}, \bibinfo {author} {\bibfnamefont {Z.}~\bibnamefont {Chen}}, \bibinfo {author} {\bibfnamefont {B.}~\bibnamefont {Chiaro}}, \bibinfo {author} {\bibfnamefont {A.}~\bibnamefont {Dunsworth}}, \bibinfo {author} {\bibfnamefont {A.}~\bibnamefont {Megrant}}, \bibinfo {author} {\bibfnamefont {P.~J.~J.}\ \bibnamefont {O’Malley}}, \bibinfo {author} {\bibfnamefont {C.}~\bibnamefont {Neill}}, \bibinfo {author} {\bibfnamefont {P.}~\bibnamefont {Roushan}}, \bibinfo {author} {\bibfnamefont {A.}~\bibnamefont {Vainsencher}}, \bibinfo {author} {\bibfnamefont {J.}~\bibnamefont
  {Wenner}}, \bibinfo {author} {\bibfnamefont {A.~N.}\ \bibnamefont {Cleland}}, \ and\ \bibinfo {author} {\bibfnamefont {J.~M.}\ \bibnamefont {Martinis}},\ }\bibfield  {title} {\enquote {\bibinfo {title} {Fast {Accurate} {State} {Measurement} with {Superconducting} {Qubits}},}\ }\href {\doibase 10.1103/PhysRevLett.112.190504} {\bibfield  {journal} {\bibinfo  {journal} {Phys. Rev. Lett.}\ }\textbf {\bibinfo {volume} {112}},\ \bibinfo {pages} {190504} (\bibinfo {year} {2014})}\BibitemShut {NoStop}%
\bibitem [{\citenamefont {Walter}\ \emph {et~al.}(2017)\citenamefont {Walter}, \citenamefont {Kurpiers}, \citenamefont {Gasparinetti}, \citenamefont {Magnard}, \citenamefont {Potočnik}, \citenamefont {Salathé}, \citenamefont {Pechal}, \citenamefont {Mondal}, \citenamefont {Oppliger}, \citenamefont {Eichler},\ and\ \citenamefont {Wallraff}}]{walter_rapid_2017}%
  \BibitemOpen
  \bibfield  {author} {\bibinfo {author} {\bibfnamefont {T.}~\bibnamefont {Walter}}, \bibinfo {author} {\bibfnamefont {P.}~\bibnamefont {Kurpiers}}, \bibinfo {author} {\bibfnamefont {S.}~\bibnamefont {Gasparinetti}}, \bibinfo {author} {\bibfnamefont {P.}~\bibnamefont {Magnard}}, \bibinfo {author} {\bibfnamefont {A.}~\bibnamefont {Potočnik}}, \bibinfo {author} {\bibfnamefont {Y.}~\bibnamefont {Salathé}}, \bibinfo {author} {\bibfnamefont {M.}~\bibnamefont {Pechal}}, \bibinfo {author} {\bibfnamefont {M.}~\bibnamefont {Mondal}}, \bibinfo {author} {\bibfnamefont {M.}~\bibnamefont {Oppliger}}, \bibinfo {author} {\bibfnamefont {C.}~\bibnamefont {Eichler}}, \ and\ \bibinfo {author} {\bibfnamefont {A.}~\bibnamefont {Wallraff}},\ }\bibfield  {title} {\enquote {\bibinfo {title} {Rapid {High}-{Fidelity} {Single}-{Shot} {Dispersive} {Readout} of {Superconducting} {Qubits}},}\ }\href {\doibase 10.1103/PhysRevApplied.7.054020} {\bibfield  {journal} {\bibinfo  {journal} {Phys. Rev. Applied}\ }\textbf {\bibinfo {volume}
  {7}},\ \bibinfo {pages} {054020} (\bibinfo {year} {2017})}\BibitemShut {NoStop}%
\bibitem [{\citenamefont {Koch}\ \emph {et~al.}(2007)\citenamefont {Koch}, \citenamefont {Yu}, \citenamefont {Gambetta}, \citenamefont {Houck}, \citenamefont {Schuster}, \citenamefont {Majer}, \citenamefont {Blais}, \citenamefont {Devoret}, \citenamefont {Girvin},\ and\ \citenamefont {Schoelkopf}}]{koch_charge-insensitive_2007}%
  \BibitemOpen
  \bibfield  {author} {\bibinfo {author} {\bibfnamefont {J.}~\bibnamefont {Koch}}, \bibinfo {author} {\bibfnamefont {T.~M.}\ \bibnamefont {Yu}}, \bibinfo {author} {\bibfnamefont {J.}~\bibnamefont {Gambetta}}, \bibinfo {author} {\bibfnamefont {A.~A.}\ \bibnamefont {Houck}}, \bibinfo {author} {\bibfnamefont {D.~I.}\ \bibnamefont {Schuster}}, \bibinfo {author} {\bibfnamefont {J.}~\bibnamefont {Majer}}, \bibinfo {author} {\bibfnamefont {A.}~\bibnamefont {Blais}}, \bibinfo {author} {\bibfnamefont {M.~H.}\ \bibnamefont {Devoret}}, \bibinfo {author} {\bibfnamefont {S.~M.}\ \bibnamefont {Girvin}}, \ and\ \bibinfo {author} {\bibfnamefont {R.~J.}\ \bibnamefont {Schoelkopf}},\ }\bibfield  {title} {\enquote {\bibinfo {title} {Charge-insensitive qubit design derived from the {Cooper} pair box},}\ }\href {\doibase 10.1103/PhysRevA.76.042319} {\bibfield  {journal} {\bibinfo  {journal} {Phys. Rev. A}\ }\textbf {\bibinfo {volume} {76}},\ \bibinfo {pages} {042319} (\bibinfo {year} {2007})}\BibitemShut {NoStop}%
\bibitem [{\citenamefont {Gambetta}\ \emph {et~al.}(2008)\citenamefont {Gambetta}, \citenamefont {Blais}, \citenamefont {Boissonneault}, \citenamefont {Houck}, \citenamefont {Schuster},\ and\ \citenamefont {Girvin}}]{gambetta_quantum_2008}%
  \BibitemOpen
  \bibfield  {author} {\bibinfo {author} {\bibfnamefont {J.}~\bibnamefont {Gambetta}}, \bibinfo {author} {\bibfnamefont {A.}~\bibnamefont {Blais}}, \bibinfo {author} {\bibfnamefont {M.}~\bibnamefont {Boissonneault}}, \bibinfo {author} {\bibfnamefont {A.~A.}\ \bibnamefont {Houck}}, \bibinfo {author} {\bibfnamefont {D.~I.}\ \bibnamefont {Schuster}}, \ and\ \bibinfo {author} {\bibfnamefont {S.~M.}\ \bibnamefont {Girvin}},\ }\bibfield  {title} {\enquote {\bibinfo {title} {Quantum trajectory approach to circuit {QED}: {Quantum} jumps and the {Zeno} effect},}\ }\href {\doibase 10.1103/PhysRevA.77.012112} {\bibfield  {journal} {\bibinfo  {journal} {Phys. Rev. A}\ }\textbf {\bibinfo {volume} {77}},\ \bibinfo {pages} {012112} (\bibinfo {year} {2008})}\BibitemShut {NoStop}%
\bibitem [{\citenamefont {{E. M. Purcell}}(1946)}]{e_m_purcell_spontaneous_1946}%
  \BibitemOpen
  \bibfield  {author} {\bibinfo {author} {\bibnamefont {{E. M. Purcell}}},\ }\bibfield  {title} {\enquote {\bibinfo {title} {Spontaneous {Emission} {Probabilities} at {Radio} {Frequencies},},}\ }\href {\doibase 10.1103/PhysRev.69.674.2} {\bibfield  {journal} {\bibinfo  {journal} {Phys. Rev.}\ }\textbf {\bibinfo {volume} {69}},\ \bibinfo {pages} {674--674} (\bibinfo {year} {1946})}\BibitemShut {NoStop}%
\bibitem [{\citenamefont {Reed}\ \emph {et~al.}(2010)\citenamefont {Reed}, \citenamefont {Johnson}, \citenamefont {Houck}, \citenamefont {DiCarlo}, \citenamefont {Chow}, \citenamefont {Schuster}, \citenamefont {Frunzio},\ and\ \citenamefont {Schoelkopf}}]{reed_fast_2010}%
  \BibitemOpen
  \bibfield  {author} {\bibinfo {author} {\bibfnamefont {M.~D.}\ \bibnamefont {Reed}}, \bibinfo {author} {\bibfnamefont {B.~R.}\ \bibnamefont {Johnson}}, \bibinfo {author} {\bibfnamefont {A.~A.}\ \bibnamefont {Houck}}, \bibinfo {author} {\bibfnamefont {L.}~\bibnamefont {DiCarlo}}, \bibinfo {author} {\bibfnamefont {J.~M.}\ \bibnamefont {Chow}}, \bibinfo {author} {\bibfnamefont {D.~I.}\ \bibnamefont {Schuster}}, \bibinfo {author} {\bibfnamefont {L.}~\bibnamefont {Frunzio}}, \ and\ \bibinfo {author} {\bibfnamefont {R.~J.}\ \bibnamefont {Schoelkopf}},\ }\bibfield  {title} {\enquote {\bibinfo {title} {Fast reset and suppressing spontaneous emission of a superconducting qubit},}\ }\href {\doibase 10.1063/1.3435463} {\bibfield  {journal} {\bibinfo  {journal} {Applied Physics Letters}\ }\textbf {\bibinfo {volume} {96}},\ \bibinfo {pages} {203110} (\bibinfo {year} {2010})}\BibitemShut {NoStop}%
\bibitem [{\citenamefont {Gard}\ \emph {et~al.}(2019)\citenamefont {Gard}, \citenamefont {Jacobs}, \citenamefont {Aumentado},\ and\ \citenamefont {Simmonds}}]{gard_fast_2019}%
  \BibitemOpen
  \bibfield  {author} {\bibinfo {author} {\bibfnamefont {B.~T.}\ \bibnamefont {Gard}}, \bibinfo {author} {\bibfnamefont {K.}~\bibnamefont {Jacobs}}, \bibinfo {author} {\bibfnamefont {J.}~\bibnamefont {Aumentado}}, \ and\ \bibinfo {author} {\bibfnamefont {R.~W.}\ \bibnamefont {Simmonds}},\ }\href {http://arxiv.org/abs/1809.02597} {\enquote {\bibinfo {title} {Fast, {High}-{Fidelity}, {Quantum} {Non}-demolition {Readout} of a {Superconducting} {Qubit} {Using} a {Transverse} {Coupling}},}\ } (\bibinfo {year} {2019})\BibitemShut {NoStop}%
\bibitem [{\citenamefont {Didier}, \citenamefont {Bourassa},\ and\ \citenamefont {Blais}(2015)}]{didier_fast_2015}%
  \BibitemOpen
  \bibfield  {author} {\bibinfo {author} {\bibfnamefont {N.}~\bibnamefont {Didier}}, \bibinfo {author} {\bibfnamefont {J.}~\bibnamefont {Bourassa}}, \ and\ \bibinfo {author} {\bibfnamefont {A.}~\bibnamefont {Blais}},\ }\bibfield  {title} {\enquote {\bibinfo {title} {Fast {Quantum} {Nondemolition} {Readout} by {Parametric} {Modulation} of {Longitudinal} {Qubit}-{Oscillator} {Interaction}},}\ }\href {\doibase 10.1103/PhysRevLett.115.203601} {\bibfield  {journal} {\bibinfo  {journal} {Phys. Rev. Lett.}\ }\textbf {\bibinfo {volume} {115}},\ \bibinfo {pages} {203601} (\bibinfo {year} {2015})}\BibitemShut {NoStop}%
\bibitem [{\citenamefont {Touzard}\ \emph {et~al.}(2019)\citenamefont {Touzard}, \citenamefont {Kou}, \citenamefont {Frattini}, \citenamefont {Sivak}, \citenamefont {Puri}, \citenamefont {Grimm}, \citenamefont {Frunzio}, \citenamefont {Shankar},\ and\ \citenamefont {Devoret}}]{touzard_gated_2019}%
  \BibitemOpen
  \bibfield  {author} {\bibinfo {author} {\bibfnamefont {S.}~\bibnamefont {Touzard}}, \bibinfo {author} {\bibfnamefont {A.}~\bibnamefont {Kou}}, \bibinfo {author} {\bibfnamefont {N.~E.}\ \bibnamefont {Frattini}}, \bibinfo {author} {\bibfnamefont {V.~V.}\ \bibnamefont {Sivak}}, \bibinfo {author} {\bibfnamefont {S.}~\bibnamefont {Puri}}, \bibinfo {author} {\bibfnamefont {A.}~\bibnamefont {Grimm}}, \bibinfo {author} {\bibfnamefont {L.}~\bibnamefont {Frunzio}}, \bibinfo {author} {\bibfnamefont {S.}~\bibnamefont {Shankar}}, \ and\ \bibinfo {author} {\bibfnamefont {M.~H.}\ \bibnamefont {Devoret}},\ }\bibfield  {title} {\enquote {\bibinfo {title} {Gated {Conditional} {Displacement} {Readout} of {Superconducting} {Qubits}},}\ }\href {\doibase 10.1103/PhysRevLett.122.080502} {\bibfield  {journal} {\bibinfo  {journal} {Phys. Rev. Lett.}\ }\textbf {\bibinfo {volume} {122}},\ \bibinfo {pages} {080502} (\bibinfo {year} {2019})}\BibitemShut {NoStop}%
\bibitem [{\citenamefont {Bergeal}\ \emph {et~al.}(2010)\citenamefont {Bergeal}, \citenamefont {Vijay}, \citenamefont {Manucharyan}, \citenamefont {Siddiqi}, \citenamefont {Schoelkopf}, \citenamefont {Girvin},\ and\ \citenamefont {Devoret}}]{bergeal_analog_2010}%
  \BibitemOpen
  \bibfield  {author} {\bibinfo {author} {\bibfnamefont {N.}~\bibnamefont {Bergeal}}, \bibinfo {author} {\bibfnamefont {R.}~\bibnamefont {Vijay}}, \bibinfo {author} {\bibfnamefont {V.~E.}\ \bibnamefont {Manucharyan}}, \bibinfo {author} {\bibfnamefont {I.}~\bibnamefont {Siddiqi}}, \bibinfo {author} {\bibfnamefont {R.~J.}\ \bibnamefont {Schoelkopf}}, \bibinfo {author} {\bibfnamefont {S.~M.}\ \bibnamefont {Girvin}}, \ and\ \bibinfo {author} {\bibfnamefont {M.~H.}\ \bibnamefont {Devoret}},\ }\bibfield  {title} {\enquote {\bibinfo {title} {Analog information processing at the quantum limit with a {Josephson} ring modulator},}\ }\href {\doibase 10.1038/nphys1516} {\bibfield  {journal} {\bibinfo  {journal} {Nature Phys}\ }\textbf {\bibinfo {volume} {6}},\ \bibinfo {pages} {296--302} (\bibinfo {year} {2010})}\BibitemShut {NoStop}%
\bibitem [{\citenamefont {Roy}\ \emph {et~al.}(2017)\citenamefont {Roy}, \citenamefont {Kundu}, \citenamefont {Chand}, \citenamefont {Hazra}, \citenamefont {Nehra}, \citenamefont {Cosmic}, \citenamefont {Ranadive}, \citenamefont {Patankar}, \citenamefont {Damle},\ and\ \citenamefont {Vijay}}]{roy_implementation_2017}%
  \BibitemOpen
  \bibfield  {author} {\bibinfo {author} {\bibfnamefont {T.}~\bibnamefont {Roy}}, \bibinfo {author} {\bibfnamefont {S.}~\bibnamefont {Kundu}}, \bibinfo {author} {\bibfnamefont {M.}~\bibnamefont {Chand}}, \bibinfo {author} {\bibfnamefont {S.}~\bibnamefont {Hazra}}, \bibinfo {author} {\bibfnamefont {N.}~\bibnamefont {Nehra}}, \bibinfo {author} {\bibfnamefont {R.}~\bibnamefont {Cosmic}}, \bibinfo {author} {\bibfnamefont {A.}~\bibnamefont {Ranadive}}, \bibinfo {author} {\bibfnamefont {M.~P.}\ \bibnamefont {Patankar}}, \bibinfo {author} {\bibfnamefont {K.}~\bibnamefont {Damle}}, \ and\ \bibinfo {author} {\bibfnamefont {R.}~\bibnamefont {Vijay}},\ }\bibfield  {title} {\enquote {\bibinfo {title} {Implementation of {Pairwise} {Longitudinal} {Coupling} in a {Three}-{Qubit} {Superconducting} {Circuit}},}\ }\href {\doibase 10.1103/PhysRevApplied.7.054025} {\bibfield  {journal} {\bibinfo  {journal} {Phys. Rev. Applied}\ }\textbf {\bibinfo {volume} {7}},\ \bibinfo {pages} {054025} (\bibinfo {year} {2017})}\BibitemShut
  {NoStop}%
\bibitem [{\citenamefont {Pfeiffer}\ \emph {et~al.}(2024)\citenamefont {Pfeiffer}, \citenamefont {Werninghaus}, \citenamefont {Schweizer}, \citenamefont {Bruckmoser}, \citenamefont {Koch}, \citenamefont {Glaser}, \citenamefont {Huber}, \citenamefont {Bunch}, \citenamefont {Haslbeck}, \citenamefont {Knudsen}, \citenamefont {Krylov}, \citenamefont {Liegener}, \citenamefont {Marx}, \citenamefont {Richard}, \citenamefont {Romeiro}, \citenamefont {Roy}, \citenamefont {Schirk}, \citenamefont {Schneider}, \citenamefont {Singh}, \citenamefont {Södergren}, \citenamefont {Tsitsilin}, \citenamefont {Wallner}, \citenamefont {Riofrío},\ and\ \citenamefont {Filipp}}]{pfeiffer_efficient_2024}%
  \BibitemOpen
  \bibfield  {author} {\bibinfo {author} {\bibfnamefont {F.}~\bibnamefont {Pfeiffer}}, \bibinfo {author} {\bibfnamefont {M.}~\bibnamefont {Werninghaus}}, \bibinfo {author} {\bibfnamefont {C.}~\bibnamefont {Schweizer}}, \bibinfo {author} {\bibfnamefont {N.}~\bibnamefont {Bruckmoser}}, \bibinfo {author} {\bibfnamefont {L.}~\bibnamefont {Koch}}, \bibinfo {author} {\bibfnamefont {N.}~\bibnamefont {Glaser}}, \bibinfo {author} {\bibfnamefont {G.}~\bibnamefont {Huber}}, \bibinfo {author} {\bibfnamefont {D.}~\bibnamefont {Bunch}}, \bibinfo {author} {\bibfnamefont {F.}~\bibnamefont {Haslbeck}}, \bibinfo {author} {\bibfnamefont {M.}~\bibnamefont {Knudsen}}, \bibinfo {author} {\bibfnamefont {G.}~\bibnamefont {Krylov}}, \bibinfo {author} {\bibfnamefont {K.}~\bibnamefont {Liegener}}, \bibinfo {author} {\bibfnamefont {A.}~\bibnamefont {Marx}}, \bibinfo {author} {\bibfnamefont {L.}~\bibnamefont {Richard}}, \bibinfo {author} {\bibfnamefont {J.}~\bibnamefont {Romeiro}}, \bibinfo {author} {\bibfnamefont {F.}~\bibnamefont
  {Roy}}, \bibinfo {author} {\bibfnamefont {J.}~\bibnamefont {Schirk}}, \bibinfo {author} {\bibfnamefont {C.}~\bibnamefont {Schneider}}, \bibinfo {author} {\bibfnamefont {M.}~\bibnamefont {Singh}}, \bibinfo {author} {\bibfnamefont {L.}~\bibnamefont {Södergren}}, \bibinfo {author} {\bibfnamefont {I.}~\bibnamefont {Tsitsilin}}, \bibinfo {author} {\bibfnamefont {F.}~\bibnamefont {Wallner}}, \bibinfo {author} {\bibfnamefont {C.}~\bibnamefont {Riofrío}}, \ and\ \bibinfo {author} {\bibfnamefont {S.}~\bibnamefont {Filipp}},\ }\bibfield  {title} {\enquote {\bibinfo {title} {Efficient {Decoupling} of a {Nonlinear} {Qubit} {Mode} from {Its} {Environment}},}\ }\href {\doibase 10.1103/PhysRevX.14.041007} {\bibfield  {journal} {\bibinfo  {journal} {Phys. Rev. X}\ }\textbf {\bibinfo {volume} {14}},\ \bibinfo {pages} {041007} (\bibinfo {year} {2024})}\BibitemShut {NoStop}%
\bibitem [{\citenamefont {Vijay}, \citenamefont {Devoret},\ and\ \citenamefont {Siddiqi}(2009)}]{vijay_invited_2009}%
  \BibitemOpen
  \bibfield  {author} {\bibinfo {author} {\bibfnamefont {R.}~\bibnamefont {Vijay}}, \bibinfo {author} {\bibfnamefont {M.~H.}\ \bibnamefont {Devoret}}, \ and\ \bibinfo {author} {\bibfnamefont {I.}~\bibnamefont {Siddiqi}},\ }\bibfield  {title} {\enquote {\bibinfo {title} {Invited {Review} {Article}: {The} {Josephson} bifurcation amplifier},}\ }\href {\doibase 10.1063/1.3224703} {\bibfield  {journal} {\bibinfo  {journal} {Review of Scientific Instruments}\ }\textbf {\bibinfo {volume} {80}},\ \bibinfo {pages} {111101} (\bibinfo {year} {2009})}\BibitemShut {NoStop}%
\bibitem [{\citenamefont {Dassonneville}\ \emph {et~al.}(2020)\citenamefont {Dassonneville}, \citenamefont {Ramos}, \citenamefont {Milchakov}, \citenamefont {Planat}, \citenamefont {Dumur}, \citenamefont {Foroughi}, \citenamefont {Puertas}, \citenamefont {Leger}, \citenamefont {Bharadwaj}, \citenamefont {Delaforce}, \citenamefont {Naud}, \citenamefont {Hasch-Guichard}, \citenamefont {Garcia-Ripoll}, \citenamefont {Roch},\ and\ \citenamefont {Buisson}}]{dassonneville_fast_2020}%
  \BibitemOpen
  \bibfield  {author} {\bibinfo {author} {\bibfnamefont {R.}~\bibnamefont {Dassonneville}}, \bibinfo {author} {\bibfnamefont {T.}~\bibnamefont {Ramos}}, \bibinfo {author} {\bibfnamefont {V.}~\bibnamefont {Milchakov}}, \bibinfo {author} {\bibfnamefont {L.}~\bibnamefont {Planat}}, \bibinfo {author} {\bibfnamefont {E.}~\bibnamefont {Dumur}}, \bibinfo {author} {\bibfnamefont {F.}~\bibnamefont {Foroughi}}, \bibinfo {author} {\bibfnamefont {J.}~\bibnamefont {Puertas}}, \bibinfo {author} {\bibfnamefont {S.}~\bibnamefont {Leger}}, \bibinfo {author} {\bibfnamefont {K.}~\bibnamefont {Bharadwaj}}, \bibinfo {author} {\bibfnamefont {J.}~\bibnamefont {Delaforce}}, \bibinfo {author} {\bibfnamefont {C.}~\bibnamefont {Naud}}, \bibinfo {author} {\bibfnamefont {W.}~\bibnamefont {Hasch-Guichard}}, \bibinfo {author} {\bibfnamefont {J.~J.}\ \bibnamefont {Garcia-Ripoll}}, \bibinfo {author} {\bibfnamefont {N.}~\bibnamefont {Roch}}, \ and\ \bibinfo {author} {\bibfnamefont {O.}~\bibnamefont {Buisson}},\ }\bibfield  {title} {\enquote
  {\bibinfo {title} {Fast {High}-{Fidelity} {Quantum} {Nondemolition} {Qubit} {Readout} via a {Nonperturbative} {Cross}-{Kerr} {Coupling}},}\ }\href {\doibase 10.1103/PhysRevX.10.011045} {\bibfield  {journal} {\bibinfo  {journal} {Phys. Rev. X}\ }\textbf {\bibinfo {volume} {10}},\ \bibinfo {pages} {011045} (\bibinfo {year} {2020})}\BibitemShut {NoStop}%
\bibitem [{\citenamefont {Kundu}\ \emph {et~al.}(2019)\citenamefont {Kundu}, \citenamefont {Gheeraert}, \citenamefont {Hazra}, \citenamefont {Roy}, \citenamefont {Salunkhe}, \citenamefont {Patankar},\ and\ \citenamefont {Vijay}}]{kundu_multiplexed_2019}%
  \BibitemOpen
  \bibfield  {author} {\bibinfo {author} {\bibfnamefont {S.}~\bibnamefont {Kundu}}, \bibinfo {author} {\bibfnamefont {N.}~\bibnamefont {Gheeraert}}, \bibinfo {author} {\bibfnamefont {S.}~\bibnamefont {Hazra}}, \bibinfo {author} {\bibfnamefont {T.}~\bibnamefont {Roy}}, \bibinfo {author} {\bibfnamefont {K.~V.}\ \bibnamefont {Salunkhe}}, \bibinfo {author} {\bibfnamefont {M.~P.}\ \bibnamefont {Patankar}}, \ and\ \bibinfo {author} {\bibfnamefont {R.}~\bibnamefont {Vijay}},\ }\bibfield  {title} {\enquote {\bibinfo {title} {Multiplexed readout of four qubits in {3D} circuit {QED} architecture using a broadband {Josephson} parametric amplifier},}\ }\href {\doibase 10.1063/1.5089729} {\bibfield  {journal} {\bibinfo  {journal} {Applied Physics Letters}\ }\textbf {\bibinfo {volume} {114}},\ \bibinfo {pages} {172601} (\bibinfo {year} {2019})}\BibitemShut {NoStop}%
\bibitem [{\citenamefont {Chen}\ \emph {et~al.}(2023)\citenamefont {Chen}, \citenamefont {Li}, \citenamefont {Lu}, \citenamefont {Warren}, \citenamefont {Križan}, \citenamefont {Kosen}, \citenamefont {Rommel}, \citenamefont {Ahmed}, \citenamefont {Osman}, \citenamefont {Biznárová}, \citenamefont {Fadavi~Roudsari}, \citenamefont {Lienhard}, \citenamefont {Caputo}, \citenamefont {Grigoras}, \citenamefont {Grönberg}, \citenamefont {Govenius}, \citenamefont {Kockum}, \citenamefont {Delsing}, \citenamefont {Bylander},\ and\ \citenamefont {Tancredi}}]{chen_transmon_2023}%
  \BibitemOpen
  \bibfield  {author} {\bibinfo {author} {\bibfnamefont {L.}~\bibnamefont {Chen}}, \bibinfo {author} {\bibfnamefont {H.-X.}\ \bibnamefont {Li}}, \bibinfo {author} {\bibfnamefont {Y.}~\bibnamefont {Lu}}, \bibinfo {author} {\bibfnamefont {C.~W.}\ \bibnamefont {Warren}}, \bibinfo {author} {\bibfnamefont {C.~J.}\ \bibnamefont {Križan}}, \bibinfo {author} {\bibfnamefont {S.}~\bibnamefont {Kosen}}, \bibinfo {author} {\bibfnamefont {M.}~\bibnamefont {Rommel}}, \bibinfo {author} {\bibfnamefont {S.}~\bibnamefont {Ahmed}}, \bibinfo {author} {\bibfnamefont {A.}~\bibnamefont {Osman}}, \bibinfo {author} {\bibfnamefont {J.}~\bibnamefont {Biznárová}}, \bibinfo {author} {\bibfnamefont {A.}~\bibnamefont {Fadavi~Roudsari}}, \bibinfo {author} {\bibfnamefont {B.}~\bibnamefont {Lienhard}}, \bibinfo {author} {\bibfnamefont {M.}~\bibnamefont {Caputo}}, \bibinfo {author} {\bibfnamefont {K.}~\bibnamefont {Grigoras}}, \bibinfo {author} {\bibfnamefont {L.}~\bibnamefont {Grönberg}}, \bibinfo {author} {\bibfnamefont {J.}~\bibnamefont
  {Govenius}}, \bibinfo {author} {\bibfnamefont {A.~F.}\ \bibnamefont {Kockum}}, \bibinfo {author} {\bibfnamefont {P.}~\bibnamefont {Delsing}}, \bibinfo {author} {\bibfnamefont {J.}~\bibnamefont {Bylander}}, \ and\ \bibinfo {author} {\bibfnamefont {G.}~\bibnamefont {Tancredi}},\ }\bibfield  {title} {\enquote {\bibinfo {title} {Transmon qubit readout fidelity at the threshold for quantum error correction without a quantum-limited amplifier},}\ }\href {\doibase 10.1038/s41534-023-00689-6} {\bibfield  {journal} {\bibinfo  {journal} {npj Quantum Inf}\ }\textbf {\bibinfo {volume} {9}},\ \bibinfo {pages} {1--7} (\bibinfo {year} {2023})},\ \bibinfo {note} {publisher: Nature Publishing Group}\BibitemShut {NoStop}%
\bibitem [{\citenamefont {Dumas}\ \emph {et~al.}(2024)\citenamefont {Dumas}, \citenamefont {Groleau-Paré}, \citenamefont {McDonald}, \citenamefont {Muñoz-Arias}, \citenamefont {Lledó}, \citenamefont {D’Anjou},\ and\ \citenamefont {Blais}}]{dumas_measurement-induced_2024}%
  \BibitemOpen
  \bibfield  {author} {\bibinfo {author} {\bibfnamefont {M.~F.}\ \bibnamefont {Dumas}}, \bibinfo {author} {\bibfnamefont {B.}~\bibnamefont {Groleau-Paré}}, \bibinfo {author} {\bibfnamefont {A.}~\bibnamefont {McDonald}}, \bibinfo {author} {\bibfnamefont {M.~H.}\ \bibnamefont {Muñoz-Arias}}, \bibinfo {author} {\bibfnamefont {C.}~\bibnamefont {Lledó}}, \bibinfo {author} {\bibfnamefont {B.}~\bibnamefont {D’Anjou}}, \ and\ \bibinfo {author} {\bibfnamefont {A.}~\bibnamefont {Blais}},\ }\bibfield  {title} {\enquote {\bibinfo {title} {Measurement-{Induced} {Transmon} {Ionization}},}\ }\href {\doibase 10.1103/PhysRevX.14.041023} {\bibfield  {journal} {\bibinfo  {journal} {Phys. Rev. X}\ }\textbf {\bibinfo {volume} {14}},\ \bibinfo {pages} {041023} (\bibinfo {year} {2024})},\ \bibinfo {note} {publisher: American Physical Society}\BibitemShut {NoStop}%
\bibitem [{\citenamefont {Chapple}\ \emph {et~al.}(2024)\citenamefont {Chapple}, \citenamefont {McDonald}, \citenamefont {Muñoz-Arias},\ and\ \citenamefont {Blais}}]{chapple_robustness_2024}%
  \BibitemOpen
  \bibfield  {author} {\bibinfo {author} {\bibfnamefont {A.~A.}\ \bibnamefont {Chapple}}, \bibinfo {author} {\bibfnamefont {A.}~\bibnamefont {McDonald}}, \bibinfo {author} {\bibfnamefont {M.~H.}\ \bibnamefont {Muñoz-Arias}}, \ and\ \bibinfo {author} {\bibfnamefont {A.}~\bibnamefont {Blais}},\ }\href {\doibase 10.48550/arXiv.2412.07734} {\enquote {\bibinfo {title} {Robustness of longitudinal transmon readout to ionization},}\ } (\bibinfo {year} {2024}),\ \bibinfo {note} {arXiv:2412.07734 [quant-ph]}\BibitemShut {NoStop}%
\bibitem [{\citenamefont {Chapple}\ \emph {et~al.}(2025)\citenamefont {Chapple}, \citenamefont {Benhayoune-Khadraoui}, \citenamefont {Richer},\ and\ \citenamefont {Blais}}]{chapple_balanced_2025}%
  \BibitemOpen
  \bibfield  {author} {\bibinfo {author} {\bibfnamefont {A.~A.}\ \bibnamefont {Chapple}}, \bibinfo {author} {\bibfnamefont {O.}~\bibnamefont {Benhayoune-Khadraoui}}, \bibinfo {author} {\bibfnamefont {S.}~\bibnamefont {Richer}}, \ and\ \bibinfo {author} {\bibfnamefont {A.}~\bibnamefont {Blais}},\ }\href {\doibase 10.48550/arXiv.2501.09010} {\enquote {\bibinfo {title} {Balanced cross-{Kerr} coupling for superconducting qubit readout},}\ } (\bibinfo {year} {2025}),\ \bibinfo {note} {arXiv:2501.09010 [quant-ph]}\BibitemShut {NoStop}%
\bibitem [{\citenamefont {Wang}\ \emph {et~al.}(2024)\citenamefont {Wang}, \citenamefont {Liu}, \citenamefont {Chen}, \citenamefont {Du}, \citenamefont {Ying}, \citenamefont {Wang}, \citenamefont {Huo}, \citenamefont {Peng}, \citenamefont {Zhu}, \citenamefont {Chen}, \citenamefont {Lu},\ and\ \citenamefont {Pan}}]{wang_999-fidelity_2024}%
  \BibitemOpen
  \bibfield  {author} {\bibinfo {author} {\bibfnamefont {C.}~\bibnamefont {Wang}}, \bibinfo {author} {\bibfnamefont {F.-M.}\ \bibnamefont {Liu}}, \bibinfo {author} {\bibfnamefont {H.}~\bibnamefont {Chen}}, \bibinfo {author} {\bibfnamefont {Y.-F.}\ \bibnamefont {Du}}, \bibinfo {author} {\bibfnamefont {C.}~\bibnamefont {Ying}}, \bibinfo {author} {\bibfnamefont {J.-W.}\ \bibnamefont {Wang}}, \bibinfo {author} {\bibfnamefont {Y.-H.}\ \bibnamefont {Huo}}, \bibinfo {author} {\bibfnamefont {C.-Z.}\ \bibnamefont {Peng}}, \bibinfo {author} {\bibfnamefont {X.}~\bibnamefont {Zhu}}, \bibinfo {author} {\bibfnamefont {M.-C.}\ \bibnamefont {Chen}}, \bibinfo {author} {\bibfnamefont {C.-Y.}\ \bibnamefont {Lu}}, \ and\ \bibinfo {author} {\bibfnamefont {J.-W.}\ \bibnamefont {Pan}},\ }\href {\doibase 10.48550/arXiv.2412.13849} {\enquote {\bibinfo {title} {99.9\%-fidelity in measuring a superconducting qubit},}\ } (\bibinfo {year} {2024}),\ \bibinfo {note} {arXiv:2412.13849 [quant-ph]}\BibitemShut {NoStop}%
\end{thebibliography}%


\begin{thebibliography}{3}%
\makeatletter
\providecommand \@ifxundefined [1]{%
 \@ifx{#1\undefined}
}%
\providecommand \@ifnum [1]{%
 \ifnum #1\expandafter \@firstoftwo
 \else \expandafter \@secondoftwo
 \fi
}%
\providecommand \@ifx [1]{%
 \ifx #1\expandafter \@firstoftwo
 \else \expandafter \@secondoftwo
 \fi
}%
\providecommand \natexlab [1]{#1}%
\providecommand \enquote  [1]{``#1''}%
\providecommand \bibnamefont  [1]{#1}%
\providecommand \bibfnamefont [1]{#1}%
\providecommand \citenamefont [1]{#1}%
\providecommand \href@noop [0]{\@secondoftwo}%
\providecommand \href [0]{\begingroup \@sanitize@url \@href}%
\providecommand \@href[1]{\@@startlink{#1}\@@href}%
\providecommand \@@href[1]{\endgroup#1\@@endlink}%
\providecommand \@sanitize@url [0]{\catcode `\\12\catcode `\$12\catcode `\&12\catcode `\#12\catcode `\^12\catcode `\_12\catcode `\%12\relax}%
\providecommand \@@startlink[1]{}%
\providecommand \@@endlink[0]{}%
\providecommand \url  [0]{\begingroup\@sanitize@url \@url }%
\providecommand \@url [1]{\endgroup\@href {#1}{\urlprefix }}%
\providecommand \urlprefix  [0]{URL }%
\providecommand \Eprint [0]{\href }%
\providecommand \doibase [0]{http://dx.doi.org/}%
\providecommand \selectlanguage [0]{\@gobble}%
\providecommand \bibinfo  [0]{\@secondoftwo}%
\providecommand \bibfield  [0]{\@secondoftwo}%
\providecommand \translation [1]{[#1]}%
\providecommand \BibitemOpen [0]{}%
\providecommand \bibitemStop [0]{}%
\providecommand \bibitemNoStop [0]{.\EOS\space}%
\providecommand \EOS [0]{\spacefactor3000\relax}%
\providecommand \BibitemShut  [1]{\csname bibitem#1\endcsname}%
\let\auto@bib@innerbib\@empty
\bibitem [{\citenamefont {{E. M. Purcell}}(1946)}]{e_m_purcell_spontaneous_1946}%
  \BibitemOpen
  \bibfield  {author} {\bibinfo {author} {\bibnamefont {{E. M. Purcell}}},\ }\bibfield  {title} {\enquote {\bibinfo {title} {Spontaneous {Emission} {Probabilities} at {Radio} {Frequencies},},}\ }\href {\doibase 10.1103/PhysRev.69.674.2} {\bibfield  {journal} {\bibinfo  {journal} {Phys. Rev.}\ }\textbf {\bibinfo {volume} {69}},\ \bibinfo {pages} {674--674} (\bibinfo {year} {1946})}\BibitemShut {NoStop}%
\bibitem [{\citenamefont {Place}\ \emph {et~al.}(2021)\citenamefont {Place}, \citenamefont {Rodgers}, \citenamefont {Mundada}, \citenamefont {Smitham}, \citenamefont {Fitzpatrick}, \citenamefont {Leng}, \citenamefont {Premkumar}, \citenamefont {Bryon}, \citenamefont {Vrajitoarea}, \citenamefont {Sussman}, \citenamefont {Cheng}, \citenamefont {Madhavan}, \citenamefont {Babla}, \citenamefont {Le}, \citenamefont {Gang}, \citenamefont {Jäck}, \citenamefont {Gyenis}, \citenamefont {Yao}, \citenamefont {Cava}, \citenamefont {de~Leon},\ and\ \citenamefont {Houck}}]{place_new_2021}%
  \BibitemOpen
  \bibfield  {author} {\bibinfo {author} {\bibfnamefont {A.~P.~M.}\ \bibnamefont {Place}}, \bibinfo {author} {\bibfnamefont {L.~V.~H.}\ \bibnamefont {Rodgers}}, \bibinfo {author} {\bibfnamefont {P.}~\bibnamefont {Mundada}}, \bibinfo {author} {\bibfnamefont {B.~M.}\ \bibnamefont {Smitham}}, \bibinfo {author} {\bibfnamefont {M.}~\bibnamefont {Fitzpatrick}}, \bibinfo {author} {\bibfnamefont {Z.}~\bibnamefont {Leng}}, \bibinfo {author} {\bibfnamefont {A.}~\bibnamefont {Premkumar}}, \bibinfo {author} {\bibfnamefont {J.}~\bibnamefont {Bryon}}, \bibinfo {author} {\bibfnamefont {A.}~\bibnamefont {Vrajitoarea}}, \bibinfo {author} {\bibfnamefont {S.}~\bibnamefont {Sussman}}, \bibinfo {author} {\bibfnamefont {G.}~\bibnamefont {Cheng}}, \bibinfo {author} {\bibfnamefont {T.}~\bibnamefont {Madhavan}}, \bibinfo {author} {\bibfnamefont {H.~K.}\ \bibnamefont {Babla}}, \bibinfo {author} {\bibfnamefont {X.~H.}\ \bibnamefont {Le}}, \bibinfo {author} {\bibfnamefont {Y.}~\bibnamefont {Gang}}, \bibinfo {author} {\bibfnamefont
  {B.}~\bibnamefont {Jäck}}, \bibinfo {author} {\bibfnamefont {A.}~\bibnamefont {Gyenis}}, \bibinfo {author} {\bibfnamefont {N.}~\bibnamefont {Yao}}, \bibinfo {author} {\bibfnamefont {R.~J.}\ \bibnamefont {Cava}}, \bibinfo {author} {\bibfnamefont {N.~P.}\ \bibnamefont {de~Leon}}, \ and\ \bibinfo {author} {\bibfnamefont {A.~A.}\ \bibnamefont {Houck}},\ }\bibfield  {title} {\enquote {\bibinfo {title} {New material platform for superconducting transmon qubits with coherence times exceeding 0.3 milliseconds},}\ }\href {\doibase 10.1038/s41467-021-22030-5} {\bibfield  {journal} {\bibinfo  {journal} {Nat Commun}\ }\textbf {\bibinfo {volume} {12}},\ \bibinfo {pages} {1779} (\bibinfo {year} {2021})},\ \bibinfo {note} {publisher: Nature Publishing Group}\BibitemShut {NoStop}%
\bibitem [{\citenamefont {Wang}\ \emph {et~al.}(2022)\citenamefont {Wang}, \citenamefont {Li}, \citenamefont {Xu}, \citenamefont {Li}, \citenamefont {Wang}, \citenamefont {Yang}, \citenamefont {Mi}, \citenamefont {Liang}, \citenamefont {Su}, \citenamefont {Yang}, \citenamefont {Wang}, \citenamefont {Wang}, \citenamefont {Li}, \citenamefont {Chen}, \citenamefont {Li}, \citenamefont {Linghu}, \citenamefont {Han}, \citenamefont {Zhang}, \citenamefont {Feng}, \citenamefont {Song}, \citenamefont {Ma}, \citenamefont {Zhang}, \citenamefont {Wang}, \citenamefont {Zhao}, \citenamefont {Liu}, \citenamefont {Xue}, \citenamefont {Jin},\ and\ \citenamefont {Yu}}]{wang_towards_2022}%
  \BibitemOpen
  \bibfield  {author} {\bibinfo {author} {\bibfnamefont {C.}~\bibnamefont {Wang}}, \bibinfo {author} {\bibfnamefont {X.}~\bibnamefont {Li}}, \bibinfo {author} {\bibfnamefont {H.}~\bibnamefont {Xu}}, \bibinfo {author} {\bibfnamefont {Z.}~\bibnamefont {Li}}, \bibinfo {author} {\bibfnamefont {J.}~\bibnamefont {Wang}}, \bibinfo {author} {\bibfnamefont {Z.}~\bibnamefont {Yang}}, \bibinfo {author} {\bibfnamefont {Z.}~\bibnamefont {Mi}}, \bibinfo {author} {\bibfnamefont {X.}~\bibnamefont {Liang}}, \bibinfo {author} {\bibfnamefont {T.}~\bibnamefont {Su}}, \bibinfo {author} {\bibfnamefont {C.}~\bibnamefont {Yang}}, \bibinfo {author} {\bibfnamefont {G.}~\bibnamefont {Wang}}, \bibinfo {author} {\bibfnamefont {W.}~\bibnamefont {Wang}}, \bibinfo {author} {\bibfnamefont {Y.}~\bibnamefont {Li}}, \bibinfo {author} {\bibfnamefont {M.}~\bibnamefont {Chen}}, \bibinfo {author} {\bibfnamefont {C.}~\bibnamefont {Li}}, \bibinfo {author} {\bibfnamefont {K.}~\bibnamefont {Linghu}}, \bibinfo {author} {\bibfnamefont {J.}~\bibnamefont
  {Han}}, \bibinfo {author} {\bibfnamefont {Y.}~\bibnamefont {Zhang}}, \bibinfo {author} {\bibfnamefont {Y.}~\bibnamefont {Feng}}, \bibinfo {author} {\bibfnamefont {Y.}~\bibnamefont {Song}}, \bibinfo {author} {\bibfnamefont {T.}~\bibnamefont {Ma}}, \bibinfo {author} {\bibfnamefont {J.}~\bibnamefont {Zhang}}, \bibinfo {author} {\bibfnamefont {R.}~\bibnamefont {Wang}}, \bibinfo {author} {\bibfnamefont {P.}~\bibnamefont {Zhao}}, \bibinfo {author} {\bibfnamefont {W.}~\bibnamefont {Liu}}, \bibinfo {author} {\bibfnamefont {G.}~\bibnamefont {Xue}}, \bibinfo {author} {\bibfnamefont {Y.}~\bibnamefont {Jin}}, \ and\ \bibinfo {author} {\bibfnamefont {H.}~\bibnamefont {Yu}},\ }\bibfield  {title} {\enquote {\bibinfo {title} {Towards practical quantum computers: transmon qubit with a lifetime approaching 0.5 milliseconds},}\ }\href {\doibase 10.1038/s41534-021-00510-2} {\bibfield  {journal} {\bibinfo  {journal} {npj Quantum Inf}\ }\textbf {\bibinfo {volume} {8}},\ \bibinfo {pages} {1--6} (\bibinfo {year} {2022})},\
  \bibinfo {note} {publisher: Nature Publishing Group}\BibitemShut {NoStop}%
\end{thebibliography}%

\end{document}


\preprint{AIP/123-QED}

\title{Supplementary Material for "The quantromon:A qubit-resonator system with orthogonal qubit and readout modes"}

\author{Kishor V. Salunkhe}
\affiliation{Department of Condensed Matter Physics and Materials Sciences \newline Tata Institute of Fundamental Research, Homi Bhabha Road, Mumbai 400005, India }

\author{Suman Kundu}%
\affiliation{School of Science, Aalto University, P.O. Box 15100, FI-00076 Aalto, Finland.}

\author{Srijita Das}
\affiliation{Department of Condensed Matter Physics and Materials Sciences \newline Tata Institute of Fundamental Research, Homi Bhabha Road, Mumbai 400005, India }

\author{Jay Deshmukh}
\affiliation{Department of Condensed Matter Physics and Materials Sciences \newline Tata Institute of Fundamental Research, Homi Bhabha Road, Mumbai 400005, India }

\author{Meghan P. Patankar}
\affiliation{Department of Condensed Matter Physics and Materials Sciences \newline Tata Institute of Fundamental Research, Homi Bhabha Road, Mumbai 400005, India }

\author{R. Vijay}
\affiliation{Department of Condensed Matter Physics and Materials Sciences \newline Tata Institute of Fundamental Research, Homi Bhabha Road, Mumbai 400005, India }

\date{\today}
\maketitle
\newpage
\section{Quantromon Hamiltonian}
\subsection{Symmetric quantromon Hamiltonian}
In this section, we derive the Hamiltonian for the general symmetric quantromon circuit consisting of five node flux variables. Here, $E_{J}$ is junction energy and $C_{J}$ is the capacitance across the junction. The interdigital capacitor $C_{R}$ is between flux node $\Phi_{1}$ and $\Phi_{4}$. The linear inductor is split into three parts where two sections $(L_{1})$ are the same to maintain the symmetry of the circuit and the remaining section $(L_{2})$ forms the loop with the two Josephson junctions. We analyze this device at integer multiples of the flux quanta $\Phi_{ext} = n\Phi_{0}$.

\begin{figure}[htbp]
\centering
\includegraphics[scale = 0.45]{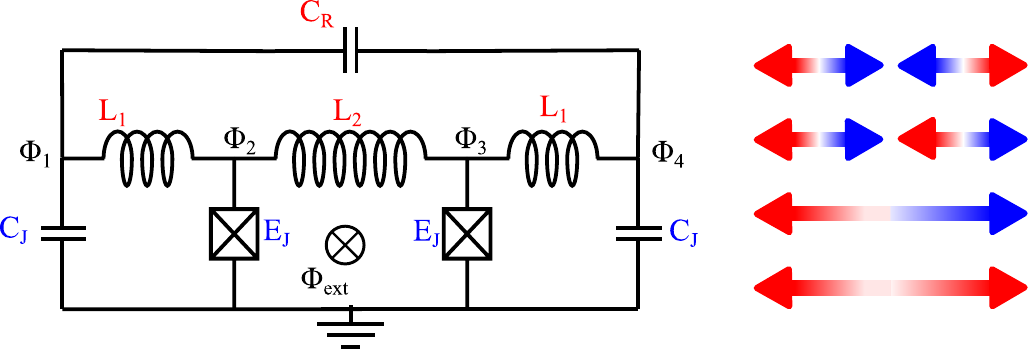}
\caption{Quantromon circuit schematic along with the mode structure; we operate at integer flux quanta of the loop formed by two Josephson junctions and inductor $L_2$.}
\label{Quantromon Schematic and mode structure}
\end{figure}

We assume that one node is grounded and the other four nodes are defined with respect to the ground, as shown in the figure. We start with writing Lagrangian in terms of generalized flux variables $\Phi_{i = 1, 2, 3, 4}$ shown in the schematic and their time derivatives. The kinetic energy $K$ of the circuit is given by,
\begin{equation}
K(\dot{\Phi_1}, \dot{\Phi_4}) =\frac{C_J}{2}\dot\Phi_{1}^{2} + \frac{C_J}{2}\dot\Phi_{4}^{2} +\frac{C_R}{2}(\dot\Phi_{1}-\dot\Phi_{4})^{2}
\end{equation}
The potential energy has the contribution from the Josephson junction ($E_J$) and the linear inductors, $L_{1}$ and $L_{2}$. The potential energy in the form of the flux variables is given by,
\begin{gather}
U(\Phi_1, \Phi_2, \Phi_3, \Phi_4) =  -E_J Cos(\frac{\Phi_2}{\Phi_0}) - E_J Cos(\frac{\Phi_3}{\Phi_0}) \\+ \frac{1}{(1-b)L_R}(\Phi_1 - \Phi_2)^2 + \frac{1}{2 b L_R}(\Phi_2 - \Phi_3)^2 \\+\frac{1}{(1-b)L_R}(\Phi_3 - \Phi_4)^2
\end{gather}
where $L_R = 2 L_1 + L_2$ is the total linear inductance and $b = L_2/L_R$ is the ratio of inductance in the loop to the total inductance.

We use the transformation given in eq[\ref{x=U*phi}] which simplifies this problem and write the Lagrangian in transformed variables $x_{i = 1, 2, 3, 4}$

\begin{equation}\label{x=U*phi}
    \begin{bmatrix}
        x_1\\
        x_2\\
        x_3\\
        x_4\\
    \end{bmatrix}
    =
    \begin{bmatrix}
        -1/2 & 1/2 & 1/2 & -1/2 \\
        -1/2 & 1/2 & -1/2 & 1/2\\
        1 & 0 & 0 & -1 \\
        1/2 & 0 & 0 & 1/2\\
    \end{bmatrix}
    \begin{bmatrix}
        \Phi_1/\Phi_0\\
        \Phi_2/\Phi_0\\
        \Phi_3/\Phi_0\\
        \Phi_4/\Phi_0\\
    \end{bmatrix}
\end{equation}

Kinetic energy and potential energy in terms of the transformed variables $\dot{x}_i$'s and $x_i$'s is

\begin{equation}
    K(\dot{x}_3, \dot{x}_4)\:=\:\left((\frac{C_R}{2} + \frac{C_J}{4})\:\dot{x_3}^2 + C_J\:\dot{x_4}^2\right)\Phi_0^2
\end{equation}

\begin{gather}\label{U(x1, x2, x3, x4)}
    U(x_1, x_2, x_3, x_4)\:=\:-E_J\cos(x_1 - x_2 - \frac{x_3}{2} + x_4) \\ - E_J \cos(x_1+x_2+\frac{x_3}{2}+x_4)\\ + E_{L_R}\left(\frac{2}{(1-b)} x_1^2+\frac{2}{b(1-b)}x_2^2
    +\frac{1}{2b}x_3^2 + \frac{2}{b} x_2x_3\right)
\end{gather}
where, $E_{L_R} = \Phi_0 ^2/L_R$. \newline

Since we don't have $\dot{x}_{1}$ and $\dot{x}_{2}$ terms in the Lagrangian our system will have constraints. These constraints can be found using the Euler-Lagrange equation by putting $\frac{d}{dt}\left( \frac{\partial L}{\partial \dot{x}_{i = 1, 2}}\right) = \frac{\partial L}{\partial x_{i = 1, 2}} = 0.$  The constraint equations for small values of $x_{i = 1, 2, 3, 4}$ are,

\begin{equation}
    \begin{aligned}
        x_1 = -\left(\frac{2E_J}{2E_J + \frac{4}{1-b} E_{L_R}}\right)x_4; \\
        x_2 = -\left(\frac{E_J + \frac{2}{b} E_{L_R}}{2E_J + \frac{4}{b(1-b)}E_{L_R}}\right)x_3
    \end{aligned}
\end{equation}

If $E_{L_R} >> E_{J}$ (this condition is true for our devices) and $0<b<1$ then $x_1 =0 $ and $x_2 = -\frac{(1-b)}{2} x_3$. We substitute these values of $x_1$ and $x_2$ to simplify our analysis and the Lagrangian contains only two variables, $x_3$ and $x_4$.

The Lagrangian is given by,
\begin{equation}
    \begin{aligned}
        L = (\frac{C_R}{2} + \frac{C_J}{4})\:\dot{x_3}^2\Phi_0^2 + C_J\:\dot{x_4}^2\Phi_0^2 + \\E_J\cos(\frac{b}{2} x_3 + x_4) + E_J\cos(\frac{b}{2} x_3 - x_4) - \frac{E_{L_R}}{2}x_3^2
    \end{aligned}
\end{equation}

Now we can write the Hamiltonian of the circuit. The capacitive energy term in the Hamiltonian is given by,

\begin{equation}
    \begin{aligned}
        H_{Cap} = \frac{{p_3}^2}{2(C_R + C_J/2)} + \frac{{p_4}^2}{2 (2 C_J)}
    \end{aligned}
\end{equation}
where $p_3 = \frac{1}{\Phi_0}\frac{\partial L}{\partial \dot{x}_3}$ and $p_4 = \frac{1}{\Phi_0}\frac{\partial L}{\partial\dot{x}_4}$ are the conjugate variables. We define the charging energies of these modes $Ec_3  =  \frac{e^2}{2(Cr + C_J/2)}$  and $ Ec_4 =  \frac{e^2}{2(2 C_J)}$ respectively and write the capacitive energy part of hamiltonian as,
\begin{equation}
    \begin{aligned}
        H_{Cap} =\frac{Ec_3}{e^2} p_3^2 + \frac{Ec_4}{e^2} p_4^2
    \end{aligned}
\end{equation}
The inductive energy term in the Hamiltonian is given by,
\begin{gather}
    H_{Ind} = -2E_J\cos(\frac{b x_3}{2})\cos(x_4) +
    \frac{E_{L_R}}{2}x_3^2
\end{gather}
We expand inductive energy up to fourth order by substituting $\cos(x) = 1 - \frac{x^2}{2} + \frac{x^4}{24}$ to get the following form

\begin{gather}
    H_{Ind} = -2E_J + \left(\frac{b^2}{4}E_J + \frac{1}{2}E_{L_R}\right)x_3^2 -\frac{b^4}{192}E_J \: x_3^4\\+ E_J \: x_4^2 -\frac{1}{12}E_J \: x_4^4 - \frac{b^2}{8}E_J \: x_3^2 x_4^2
\end{gather}
where we have quadratic and quartic term for each mode and coupling $x_3^2x_4^2$.
The full hamiltonian of the circuit is $H_{Cap} + H_{Ind}$,
\begin{gather}
    H = -2E_J + \frac{Ec_3}{e^2} p_3^2 +  \left(\frac{b^2}{4}E_J + \frac{1}{2}E_{L_R}\right)x_3^2 -\frac{b^4}{192}E_J \: x_3^4\\+ \frac{Ec_4}{e^2} p_4^2 + E_J x_4^2 -\frac{E_J}{12} x_4^4 - \frac{b^2}{8}E_J \: x_3^2 x_4^2
\end{gather}
This Hamiltonian is of two coupled anharmonic oscillators. These oscillators are the two orthogonal modes of our circuit. The anharmonicity of the mode associated with $x_3$ is diluted and can be treated as a harmonic oscillator ($Ec_3 << Ec_4$), whereas the other one is the transmon-like mode with effective capacitance $2C_J$ and effective Josephson junction $2E_J$. The coupling between the modes is via cross-Kerr interaction $x_3^2x_4^2$. From now on we will address the mode associated with $x_3$ as resonator mode and other as a qubit mode. We will change the subscripts in notation from $(3,4)$ to $(r, q)$.
\par We quantize the Hamiltonian by introducing the raising and lowering operators. The position and momenta of qubit and resonator are given as

\begin{align}
    p_r &= -i \sqrt{\frac{\hbar}{2 Z_r}} (\hat{a_r} - \hat{a_r}^\dagger) ; \:\:\:  x_r = \sqrt{\frac{\hbar Z_r}{2}} (\hat{a_r} + \hat{a_r}^\dagger) \\
    p_q &= -i \sqrt{\frac{\hbar}{2 Z_q}} (\hat{a_q} - \hat{a_q}^\dagger) ; \:\:\:  x_q = \sqrt{\frac{\hbar Z_q}{2}} (\hat{a_q} + \hat{a_q}^\dagger)
\end{align}

where the uncoupled mode frequencies and impedances are defined as follows,

\begin{align}
    \omega_r &= \frac{\sqrt{8E_{J_R}E_{C_R}}}{\hbar};\quad Z_r = \frac{\hbar}{e^2}\sqrt{\frac{E_{C_R}}{E_{J_R}}};\quad E_{J_R} = \frac{b^2}{2}E_J + E_{L_R} \\
    \omega_q &= \frac{\sqrt{8E_{J_Q}E_{C_Q}}}{\hbar};\quad Z_q = \frac{\hbar}{e^2}\sqrt{\frac{E_{C_Q}}{E_{J_Q}}};\quad E_{J_Q} = 2E_J
\end{align}

We use RWA approximation to keep the terms which conserves the number of excitaiton in the qubit-resonator system. The terms corresponding to the exchange of two excitations are also ignored here assuming large detuning between qubit and resonator.
We write the Hamiltonian which resembles the Jaynes-Cummings Hamiltonian in the dispersive limit.
\begin{equation}
    \frac{H_{eff}}{\hbar} = \tilde{\omega}_q \hat{n}_q - \frac{\alpha_q}{2} \hat{n}_q(\hat{n}_q-1)+\tilde{\omega}_r \hat{n}_r - 2 \chi \hat{n}_r\hat{n}_q
\end{equation}
where,
\begin{align}
    \tilde{\omega}_q &= \left(\sqrt{8E_{J_Q}E_{C_Q}} - E_{C_Q} - \frac{b^2}{2}\sqrt{{\frac{E_{C_R} E_{C_Q}}{ \frac{b^2}{2} + \frac{E_{L_R}}{E_J}}}}\right)/\hbar\\
    \alpha_q &= E_{C_Q}/\hbar\\
    \tilde{\omega}_r &= \left(\sqrt{8E_{J_R}E_{C_R}}-\frac{b^2}{2}\sqrt{{\frac{E_{C_R} E_{C_Q}}{ \frac{b^2}{2} + \frac{E_{L_R}}{E_J}}}}\right)/\hbar\\
    \chi &= \frac{b^2}{2\sqrt{2}\hbar}\sqrt{{\frac{E_{C_R} E_{C_Q}}{ \frac{b^2}{2} + \frac{E_{L_R}}{E_J}}}}
\end{align}
The $\tilde{\omega}_q$ and $\tilde{\omega}_r$ are the renormalized frequencies of the qubit mode and the resonator mode. The anharmonicity of the qubit mode is given by $\alpha_q$. The anharmonicity in the resonator mode $\alpha_r$ is ignored here as $E_{L_R}>>E_J$. The dispersive shift ($2\chi$) is dependent on the ratio of the ratio $b$. The dispersive shift is zero when no part of the linear inductor is inside the quantromon loop. The dispersive shift is maximum when the entire inductor is a part of the quantromon loop. In the quantromon geometry, we can make the dispersive shift very large without compromising on the Purcell effect for fixed qubit and resonator parameters.

\subsection{Asymmetric quantromon}
The two junctions can be asymmetric due to the fabrication uncertainties. In this section, we analyze the effect of asymmetry between two junctions. We write the junction energies as follows,

\begin{align*}
    E_{J1} &= \frac{1 + d_J}{2}E_{J\Sigma}\\
    E_{J2} &= \frac{1 - d_J}{2}E_{J\Sigma}
\end{align*}
where the $E_{J\Sigma} = E_{J1} + E_{J2}$ and $d_J = \frac{E_{J1} - E_{J2}}{E_{J\Sigma}}$ is the asymmetry between the two juctions. The junction asymmetry will modify the inductive energy term in the quantromon Hamiltonian to

\begin{gather}
    H_{Ind} = -E_{J\Sigma} + \frac{1}{2}\left(\frac{b^2}{4} E_{J\Sigma} + E_{L_R}\right)x_3^2 -\frac{b^2}{384}E_{J\Sigma} x_3^4\\\frac{E_{J\Sigma}}{2} x_4^2 -\frac{E_{J\Sigma}}{24} x_4^4 -\frac{b^2}{16}E_{J\Sigma} x_3^2 x_4^2 - d_J\frac{b}{2} E_{J\Sigma}x_3x_4
\end{gather}
where the inductive energies of modes are given by,
\begin{align*}
    E_{J_R} &= \frac{b^2}{4}E_{J\Sigma} + E_{L_R};\quad E_{J_Q} = E_{J\Sigma}
\end{align*}

We follow the same procedure to write the Hamiltonian in quantized form as,

\begin{align}
     \frac{H_{eff}}{\hbar} = \tilde{\omega}_q \hat{n}_q - \frac{\alpha_q}{2} \hat{n}_q(\hat{n}_q-1)+\tilde{\omega}_r \hat{n}_r - 2 \chi \hat{n}_r\hat{n}_q \\ + g_{asymm}(\hat{a_r} + \hat{a_r}^\dagger)(\hat{a_q} + \hat{a_q}^\dagger)
\end{align}

where, we have an extra term that due to the residual transverse coupling due to asymmetry in the junctions and the strength of the coupling is give by
\begin{equation}
   g_{asymm} = -d_J\sqrt{2\chi (E_{J\Sigma}/\hbar)}
\end{equation}

where $\chi$ is dispersive shift due to cross-kerr coupling. We write the interaction term due to junction asymmetry in the dispersive limit to get the hamiltonian,

\begin{equation}
    \frac{H_{eff}}{\hbar} = \tilde{\omega}_q \hat{n}_q - \frac{\alpha_q}{2} \hat{n}_q(\hat{n}_q-1)+\tilde{\omega}_r \hat{n}_r - 2 \tilde{\chi} \hat{n}_r\hat{n}_q
\end{equation}

where,
\begin{equation}\label{chi_total_asymm}
    \tilde{\chi} = \left(1 + 2d_J^2\frac{E_{J\Sigma}}{\hbar}\frac{\alpha_q}{\Delta(\Delta+\alpha_q)}\right) \chi
\end{equation}

the $\tilde{\chi}$ is the total dispersive shift, which contains dispersive shift due to cross-kerr coupling and residual transverse coupling. The contribution due to the transverse coupling will dominate if the qubit resonator detuning is small.

\subsection{Purcell $T_1$ due to asymmetry}
To estimate the Purcell decay \cite{e_m_purcell_spontaneous_1946} due to asymmetry, we need to know the transverse coupling due to the asymmetry and qubit-cavity detuning. The dispersive shift due to the cross-Kerr coupling is calculated by inverting the eq\ref{chi_total_asymm} and using the experimentally measured dispersive shift ($\tilde{\chi}$), qubit-cavity detunig ($\Delta$), qubit anharmonicity ($\alpha_q$), qubit inductive energy ($E_{J\Sigma}$) and estimated asymmetry ($d_j$). We use the dispersive shift due to the cross-Kerr interaction and junction asymmetry to calculate the $g_{asymm} = -d_J\sqrt{2\chi (E_{J\Sigma}/\hbar)}$. The Purcell $T_1$, $T_{asymm}$ is given by,
\begin{equation}
    T_{asymm} = \frac{1}{\kappa}\frac{g_{asymm}^2}{\Delta^2}
\end{equation}
 where $\kappa$ is the line-width of the cavity.

\begin{table}[t]
\centering
\begin{tabular}{ccc}
\hline
Parameter & Theory fit & FE Simulation \\
\hline
$L_J$ (nH) & 8.2 & --  \\
$C_J$ (fF)  & 56.88 & 50.23 \\
$L_R$ (nH) & 0.546 & 0.503  \\
$C_R$ (fF) & 781.8 & 713.9  \\
$L_2/L_R$ & 0.405 & 0.324 \\
\hline
\end{tabular}
\caption{The parameters used for the theory fit and obtained from the finite element simulation  }
\label{Dispersive Expt params}
\end{table}

\section{Fabrication details}
The samples used in the main results of the paper are fabricated using two different methods. Sample A is fabricated on a silicon substrate using the standard Dolan bridge technique where $20$~nm thin film is deposited in the first evaporation and $50$~nm thin film in the second evaporation. We grow aluminum oxide naturally on top of the first film by exposing it to oxygen. The mask for the evaporation is patterned in the bilayer, $\sim450$~nm of EL9 is the bottom layer and $\sim200$~nm of ARP6200 is the top layer.

Samples B and C are tantalum-based \cite{place_new_2021,wang_towards_2022} and fabricated on the tantalum sputtered sapphire chip. We use ARN7520 resits after precleaning the chip with Ar-plasma to create a mask by exposing the large features of three big pads P1, P2, and P3. We use a dry etch of the tantalum from the unwanted area using SF6 plasma. The resist usually hardens after dry etch and is difficult to remove by prescribed remover so, we do a short dip in Transene which pills of the resits. We proceed with the fabrication method used for Sample A for the rest of the process. We make the interdigitated capacitors in aluminum instead of tantalum since our tantalum etching process is not yet optimized for the small resolution needed.

All samples are cleaned in Ar-plasma to remove the resistant residues before loading them in the evaporator. We do in-situ Ar-plasma cleaning before the evaporation of tantalum samples to remove the oxide.

\section{Device paramters}

The parameters used for the theory fit in the main paper are presented in the table \ref{Dispersive Expt params}. The parameters $L_J$ and $C_J$ are extracted using experimentally measured qubit frequency and anharmonicity. Extracting parameters for the resonator is tricky as the data set is not sufficient. We fix the inductor ratio $b$ and extract the $L_R$ and $C_R$ from the measured resonator frequency and dispersive shift. Here, we assume an asymmetry of $4.5\%$ to calculate the contribution from the transverse dispersive shift. The values used in the fit are close to the values obtained from the simulation using HFSS.

\section{Single shot readout fidelity}
\begin{table}[t]
\centering
\begin{tabular}{cccc}
\hline
Sample Name & Readout Fidelity & Int.Time($\mu$s) & Material \\
\hline
QuantroTa01 A & 98.3\% & 1.8 & Tantalum \\
QuantroTa01 B & 97.5\% & 1.2 & Tantalum \\
QuantroTa01 C & 97.6\% & 1.4 & Tantalum \\
Quantro04  B & 97.8\% & 0.8 & Aluminum \\
\hline
\end{tabular}
\caption{The readout fidelity numbers for the device with different choices of material are shown. We get fidelities > 97.5\%  for all the devices mentioned in the table.}
\label{Fidelity table}
\end{table}
The results for the readout fidelity for the different quantromon devices is mentioned in the table \ref{Fidelity table}. We consistently get high fidelity (>$97.5\%$) for the devices fabricated using tantalum and aluminum. The integration time and readout power are optimized to get high fidelity. We used uniform integration weights during the integration.

\section{Wave-guide details}

\begin{figure}[htbp]
\centering
\includegraphics[scale = 0.3]{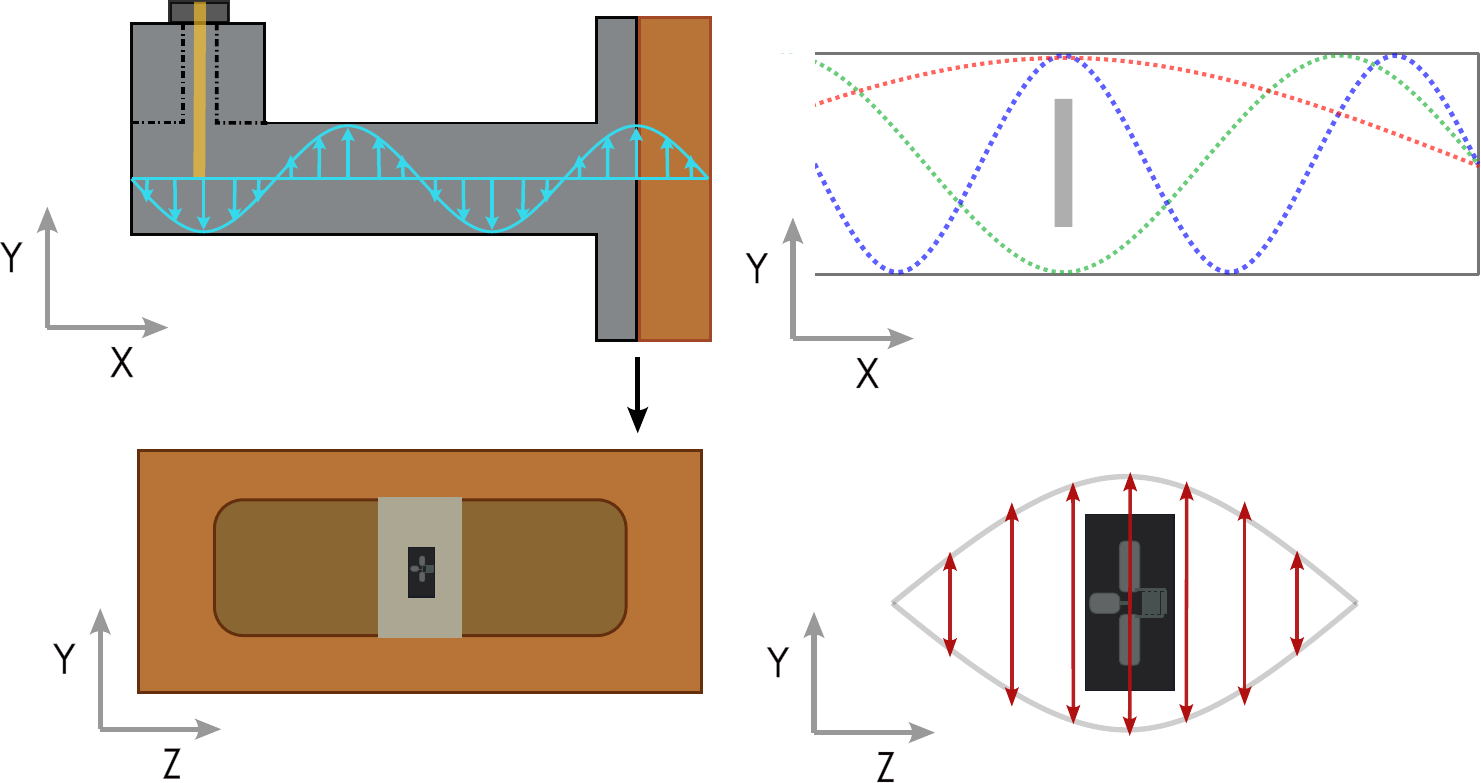}
\caption{\textbf{Waveguide:} (a) The Schematic of the waveguide. We use the launch single coming from coax into the waveguide.}
\label{Waveguide}
\end{figure}

All quantromon devices are measured using a rectangular waveguide of dimensions $1.75\times3.5\times8.2$~cm made in two parts.  We use aluminum (6061 alloy) for the waveguide to ensure low loss and a copper termination to enable flux tunability. The signal is launched into the waveguide from a coaxial port coupled to the waveguide at one end. The cutoff frequency ($f_c$) is set by the lowest mode of the waveguide, $ f_c\sim 4.3$ GHz. The chip position ($12$ mm from the other end) is decided such that the amplitude for the frequencies($\approx 7.0$ GHz -$7.5$ GHz) is maximum, ensuring the maximum coupling for the typical range for our readout resonators. The electric field of the qubit is orthogonal and does not suffer the Purcell decay via waveguide if the qubit frequency is less than $\sim 9.5$ GHz.The external bandwidth of the readout mode is set by the dimensions of the antennas (pads P1 and P3 in Fig.~1 of manuscript) and the chip's position in the waveguide. The qubit and the resonator are excited from the same port; this is possible since there is always a small asymmetry in the qubit junctions which allows a small but sufficient transverse coupling for qubit-control.

\section{Cryogenic Microwave Setup}
We use a standard reflection setup for all the measurements mentioned in the main paper. The input line has a total 60dB attenuation (20dB at 4K and 40dB at the base plate). The eccosorb (IR) and low pass filters are used on both the input and output line. The output signal is amplified by at the 4K stage and room temperature stage using low-noise amplifiers.

\bibliography{Quantromon_suppl_reference}